\documentclass[aps,prl,twocolumn,nofootinbib,superscriptaddress]{revtex4-2}
\pdfoutput=1
\usepackage{graphicx}  
\usepackage{dcolumn}   
\usepackage{bm}        
\usepackage{amsmath,amssymb,amsxtra,bm,epsfig}   
\usepackage{float}
\usepackage[dvipsnames]{xcolor}
\usepackage{xspace}
\usepackage{hyperref}
\usepackage{cancel,soul,ulem}
\hypersetup{
	colorlinks=true,
	linkcolor=red,
	filecolor=magenta,
	urlcolor=ForestGreen,
	citecolor=RoyalBlue}

\usepackage[]{caption}
\newcommand{\bmt}{\begin{pmatrix}}
	\newcommand{\emt}{\end{pmatrix}}
\newcommand{\ba}{\begin{array}{c}}
	\newcommand{\ea}{\end{array}}
\newcommand{\be}{\begin{equation}}
	\newcommand{\ee}{\end{equation}}
\newcommand{\bea}{\begin{eqnarray}}
	\newcommand{\eea}{\end{eqnarray}}

\newcommand{\bi}{\begin{itemize}}
	\newcommand{\ei}{\end{itemize}}
\newcommand{\baz}{\begin{array}{cc}}
\newcommand{\besub}{\begin{subequations}}
\newcommand{\eesub}{\end{subequations}}

\begin{document}
	\title{Effects of Reheating on Charged Lepton Yukawa Equilibration and Leptogenesis}
	
	\author{Arghyajit Datta}
	\email{arghyad053@gmail.com}
	\affiliation{Department of Physics, Kyungpook National University, Daegu 41566, Republic of Korea}
	\affiliation{Center for Precision Neutrino Research, Chonnam National University, Gwangju 61186, Republic of Korea }
	\affiliation{Department of Physics, Indian Institute of Technology Guwahati, Assam 781039, India}		
	
	\author{Rishav Roshan}
	\email{rishav.roshan@gmail.com}
	\affiliation{School of Physics and Astronomy, University of Southampton,\\ Southampton, SO17 1BJ, U.K.}
	
	\author{Arunansu Sil}
	\email{asil@iitg.ac.in}
	\affiliation{Department of Physics, Indian Institute of Technology Guwahati, Assam 781039, India}

		\begin{abstract}
			We show that the process of non-instantaneous reheating during the post-inflationary period can have a sizable impact on the charged lepton Yukawa equilibration temperature in the early Universe. This suggests relooking the effects of lepton flavors in the leptogenesis scenario where the production and decay of right-handed neutrinos take place within this prolonged era of reheating. We find this observation has the potential to shift the flavor regime(s) of leptogenesis compared to the standard thermal scenario.
			
		\end{abstract}

		\maketitle
		
The observed dominance of matter over antimatter in our Universe~\cite{Planck:2018vyg} requires 
	an explanation where such an asymmetry is expected to be generated in a post-inflationary phase as otherwise, 
	any pre-inflationary asymmetry, if present, would be erased due to the exponential expansion of the Universe 
	during inflation. Among many possibilities, production of lepton asymmetry resulting from out-of-equilibrium decay of heavy right-handed neutrinos (RHN) $N$ into Standard Model (SM) lepton ($\ell_L^T = (\nu_L, e_L)$) and Higgs ($H$) doublets, called $\it{leptogenesis}$~\cite{Fukugita:1986hr,Luty:1992un,Pilaftsis:1997jf}, remains an attractive one. The obvious reason behind it lies in its close proximity with neutrino mass ($m_{\nu}$) generation mechanism via type-I seesaw~\cite{Minkowski:1977sc,Yanagida:1979as,Yanagida:1979gs,GellMann:1980vs,Mohapatra:1979ia,Schechter:1980gr,Schechter:1981cv,Datta:2021elq} where the Lagrangian is given by, 
\begin{align}
	-\mathcal{L}= \overline{\ell}_{L_\alpha} (Y_{\nu})_{\alpha i} \tilde{H} N_{i}+ \frac{1}{2}  \overline{N_{i}^c}(M_{R})_{i} N_i+ h.c.,
\end{align}
(in the charged lepton diagonal basis) with $\alpha = e, \mu, \tau$ and $i=1,2,3$. Such a lepton asymmetry then reprocesses itself in terms of baryon asymmetry by $B+L$ violating 
sphaleron transitions~\cite{Kuzmin:1985mm,Bento:2003jv,DOnofrio:2014rug} (while conserving $B-L$) which remains in equilibrium from a temperature $T_s \sim 10^{12}$ GeV down to electroweak (EW) scale $T_{EW} \sim$ 100 GeV \cite{Arnold:1996dy,Bodeker:1998hm,Arnold:1998cy,Bento:2003jv} in a radiation dominated Universe. 

In the standard thermal leptogenesis scenario, the lepton or the $B-L$ asymmetry is generated during the radiation-dominated era which is indicative of the underlying assumption that the reheating temperature ($T_{\text{RH}}$) after inflation is greater than the mass of the decaying (lightest) RHN, say $M_1$. At this end, to study the evolution of this $B-L$ asymmetry, effect of different SM interactions in terms of their respective equilibration temperatures ($T^*$) becomes very much relevant. This is due to the fact that a particle species (say $X$ with degrees of freedom $g_{X}$) involved in thermal equilibrium at a given temperature $T$ experiences chemical equilibrium conditions leading to an inter-connection between chemical potentials ($\mu_{X}$) of different species while other constraints followed from conservation laws are also enforced as outlined in the Supplemental Material. In terms of particle (antiparticle) number density 
$n_{X}$ ($n_{{\bar{X}}}$) to entropy ($s$) ratio defined by\cite{Kolb:1990vq} $Y_{\Delta_X} = n_{\Delta_{X}}/s = (n_{X} - n_{\bar{X}})/s \equiv g_{X} \frac{T^3}{6s}(\mu_{X}/T)$, the lepton asymmetry $Y_L\equiv Y_{\Delta_L}$ generated from $N$ decay becomes~\cite{Nardi:2005hs} 
\begin{equation}
	Y_{L} = \frac{n_{\Delta L}}{s} = \sum_{\alpha} (2 Y_{\Delta_{\ell_{\alpha}}} + Y_{\Delta_{e_{R_\alpha}}}),
\end{equation}
where $Y_{\Delta_{e_{R_\alpha}}}$ is the asymmetry of the right handed charged leptons, if any. This $Y_{L}$ in turn directly determine the $B-L$ asymmetry, $Y_{B-L} = -Y_{L}$ when $B = 0$. 

\begin{figure}[h]
	\includegraphics[width=0.95\linewidth]{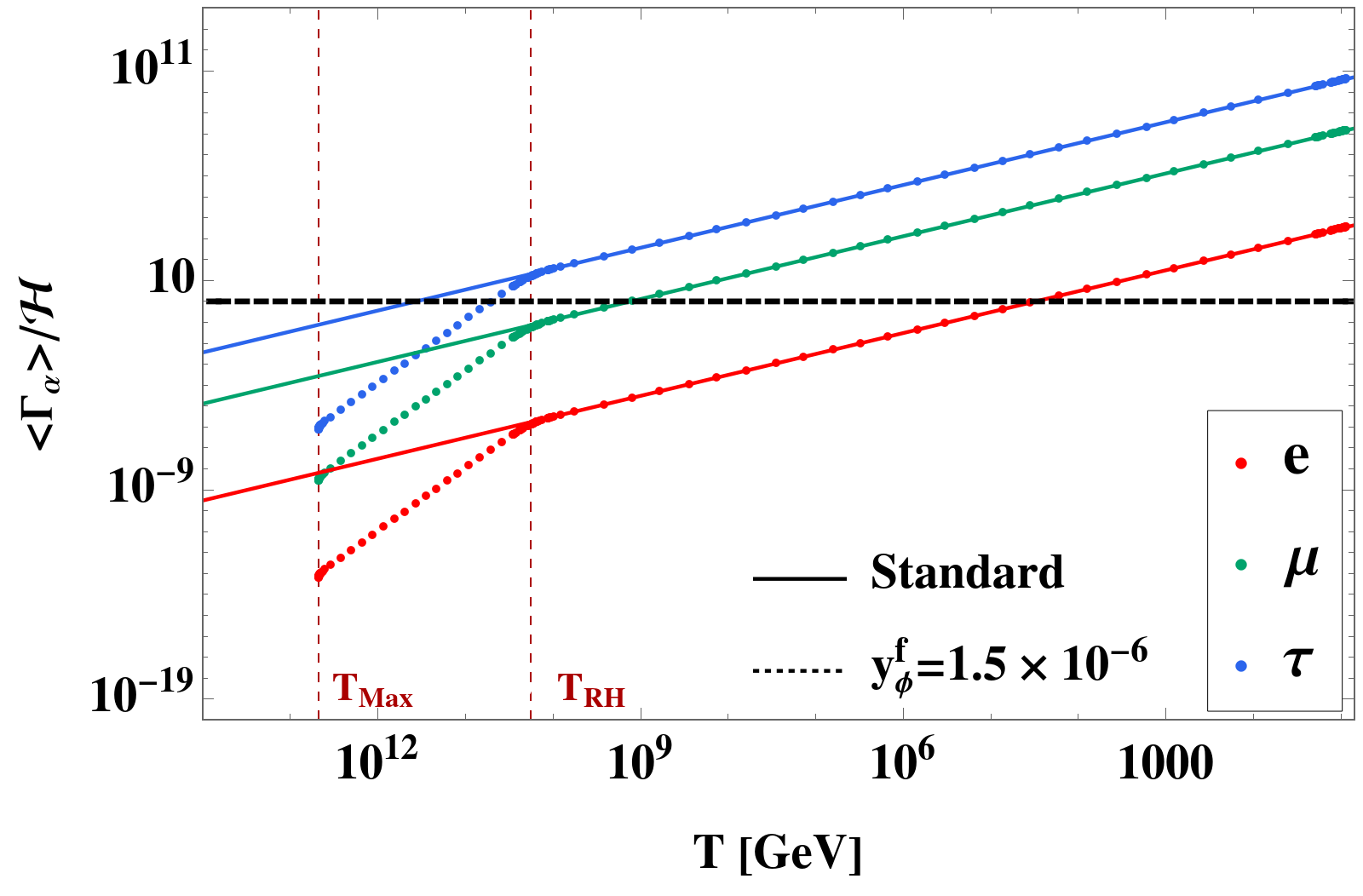}
	\caption{ Variation of $\langle\Gamma_\alpha \rangle/\mathcal{H}$ $w.r.t.$ $\rm{T}$ for standard (solid lines) and modified (dotted) scenarios.}\label{fig:a}
	\label{gamma-by-H}
\end{figure}

Note that as long as the charged lepton Yukawa ($Y_{\alpha}$) interactions do not enter equilibrium, $Y_{L}$ is confined within the asymmetry in the lepton doublets only as $Y_{\Delta_{e_{R_\alpha}}}$ = 0. The temperature at which a specific $e_{R_\alpha}$ equilibrates with corresponding lepton doublet, known as the equilibration temperature (ET), can be determined naively by comparing the Higgs boson decay (and inverse decay) rate $\langle \Gamma_{\alpha} \rangle$ with that of the Hubble expansion $\mathcal{H}$~\cite{Cline:1993vv,Cline:1993bd}. For example, the $\tau_R$ equilibration temperature in a radiation dominated (RD) Universe is obtained by solving $\langle \Gamma_{\tau} \rangle = \mathcal{H}(T)$ or equivalently 
\begin{equation}
	\frac{\pi Y^2_{\tau}}{192 \zeta(3)} \frac{m^2_h (T)}{T} = 0.066 {g_*}^{1/2}\frac{T^2}{M_P},
	\label{eq-temp}
\end{equation}
where $\mathcal{H}(T) = 0.066 {g_*}^{1/2}\frac{T^2}{M_P}$ is used ($M_P$ is the reduced Planck mass), $g_*$ represents effective number of relativistic degrees of freedom, 	
and $m_h (T)\simeq0.6T$~\cite{Weldon:1982bn, Quiros:1999jp,Senaha:2020mop} is the thermal mass of the Higgs. The $\tau_R$-ET $T^*_{0({\tau})}$ is found to be $5 \times 10^{11}$ GeV as seen from the intersection point of solid blue line (standard) and $\langle \Gamma_{\tau} \rangle/{\mathcal{H}} = 1$ dotted horizontal line of Fig. \ref{gamma-by-H}. For $\mu_R$ and $e_R$, such temperatures $T^*_{0({\mu})}$ and $T^*_{0(e)}$ are found to be $10^9$ GeV and $5 \times 10^4$ GeV respectively. Though a more refined analysis in the context of thermal field theory \cite{Garbrecht:2013bia,Bodeker:2019ajh} may slightly alter these numbers, such naive estimates of ETs remain sufficient for our purpose.

This finding plays a pivotal role behind the loss of lepton flavor coherence (fully or partially) as observed in flavor leptogenesis~\cite{Barbieri:1999ma,Nardi:2005hs,Abada:2006fw,Nardi:2006fx,Blanchet:2006be,Dev:2017trv}. Depending on which charged lepton Yukawa interaction(s) is(are) in equilibrium at a given temperature $T$, the lepton doublet state(s) ${\ell}_{L_\alpha}$ produced from the RHN-decay finds corresponding $e_{R{_\alpha}}$ in the bath to interact. As a result, the quantum coherence of the lepton doublet states gets broken, which in turn, dictates the number of orthogonal flavor alignments in that particular temperature regime.

A basic assumption behind such standard evaluation of ET is that the entire epoch happens below the reheating temperature, or in other words, a RD phase is prevalent. Though this sounds reasonable, at the same time, we can also recollect that reheating might not be an instantaneous process~\cite{Giudice:2000ex}. In its simplest version, the process of reheating begins in a matter-dominated phase where the oscillatory inflaton field ($\phi$)  decays perturbatively. The thermalization of these light decay products helps the Universe to attain quickly a maximum temperature $T_{\text{Max}}$, subsequent to which the temperature falls to $T_{\text{RH}}$ marking the commencement of RD era thereafter. During this period ($\it{i.e.}$ $T_{\text{RH}} < T < T_{\text{Max}}$), temperature varies less sharply as compared to standard scaling $T \propto a^{-1}$ ($a$ is the Friedmann–Robertson–Walker scale factor) and the Universe also expands faster than in the RD era at a given temperature. 

In this letter, we show that such non-instantaneous reheating epoch exhibits a shift in ET as 
$\mathcal{H}$ would differ from that of a purely RD era, a situation which has not been discussed elsewhere to the best of our knowledge. In view of recent finding~\cite{Garcia:2020eof} that $T_{\text{Max}}$ can be much larger than $T_{\text{RH}}$, this study can be even more relevant. We further elaborate here for the first time\footnote{ Earlier works on flavor leptogenesis during reheating \cite{Antusch:2006gy,Blanchet:2006be,Antusch:2007km} or within different modified cosmological scenarios\cite{Mahanta:2019sfo,Perez-Gonzalez:2020vnz} did not incorporate such shift(s) in ET(s).
} 
the impact of such shift in ET (due to the unusual relation between $T$ and $a$) on the study of flavor leptogenesis too where the production and decay of RHN take place within $T_{\text{Max}}$ and $T_{\text{RH}}$. We begin the discussion on reheating below to estimate the modified temperature regimes where the right-handed charged leptons enter equilibrium, first without introducing any RHNs, and later we incorporate RHNs so as to discuss their influence on flavored leptogenesis. 

As the era of reheating starts following the inflationary epoch, the discussion of reheating is somewhat connected to the form of inflaton potential near its minimum. To be specific, following \cite{Garcia:2020eof} we consider a generic power-law form of the inflaton potential near the minimum, $V(\phi) = \lambda \frac{|\phi|^n}{M_P^{n-4}}$ the origin of which can be traced back to T-attractor models in no-scale supergravity \cite{Kallosh:2013hoa}. For simplicity, we confine ourselves to a $n=2$ case resembling the Starobinsky type of inflationary potential~\cite{Starobinsky:1980te,Ellis:2013xoa,Khalil:2018iip} in the large field limit. Here we skip details of inflation (see \cite{Ellis:2015pla}) except the information that the magnitude of the parameter $\lambda$ is found to be $\sim 2 \times 10^{-11}$ so as to be consistent with inflationary observables like spectral index and tensor-to-scalar ratio within 95$\%$ C.L. of the Planck+BICEP2/Keck (PBK) constraints~\cite{BICEP2:2018kqh}. This leads to the effective mass of the inflaton as $m_{\phi} = (\partial ^2_{\phi} V(\phi))^{1/2} \simeq 1.5 \times 10^{13}$ GeV. 

We introduce an effective interaction between the inflaton and any specific pair of SM fermion (${f}$) anti-fermion fields as $y^f_{\phi} \phi  \bar{f} f$ which will dictate the decay of the inflaton to radiation (composed of massless SM fermion fields in the early Universe). The origin of such effective interaction can be traced back to dimension-5 operator involving SM Yukawa interactions 
such as $\frac{\alpha_{\ell}}{\Lambda}\phi \bar{\ell}_L H E_R, \quad \frac{\alpha_{u}}{\Lambda}\phi \bar{q}_L \tilde{H} u_R, \quad  \frac{\alpha_{d}}{\Lambda}\phi \bar{q}_L H d_R$, where $\Lambda$ and $\alpha_{f}$ correspond to a cut-off scale and coupling constant(s) respectively. This would initiate a decay of inflaton to Higgs and a pair of fermion anti-fermion fields where a subsequent fast decay of the Higgs into another fermion anti-fermion pair (via SM Yukawa interactions) analogously represents the decay of the inflaton into fermion anti-fermion pair. Considering all such possible final state fermion anti-fermion pairs (denoted by $g_f$ number of possibilities) couple uniformly to $\phi$, the 
total decay width of $\phi$ can be represented by $\Gamma_{\phi} \simeq g_f^2 \frac{|y^f_{\phi}|^2}{8\pi} m_{\phi}$.
Furthermore, $y^f_{\phi}$ is kept sufficiently small (quantified later) so as to exclude 
the possibility of parametric excitation of fermions \cite{Greene:1998nh} and ensure the perturbative decay 
of $\phi$. The role of reheating is therefore limited here in changing the thermal history of the Universe to clearly exhibit the impact on charged lepton Yukawa ET.

\begin{figure}[h]
	\centering
	{\includegraphics[width=1\linewidth]{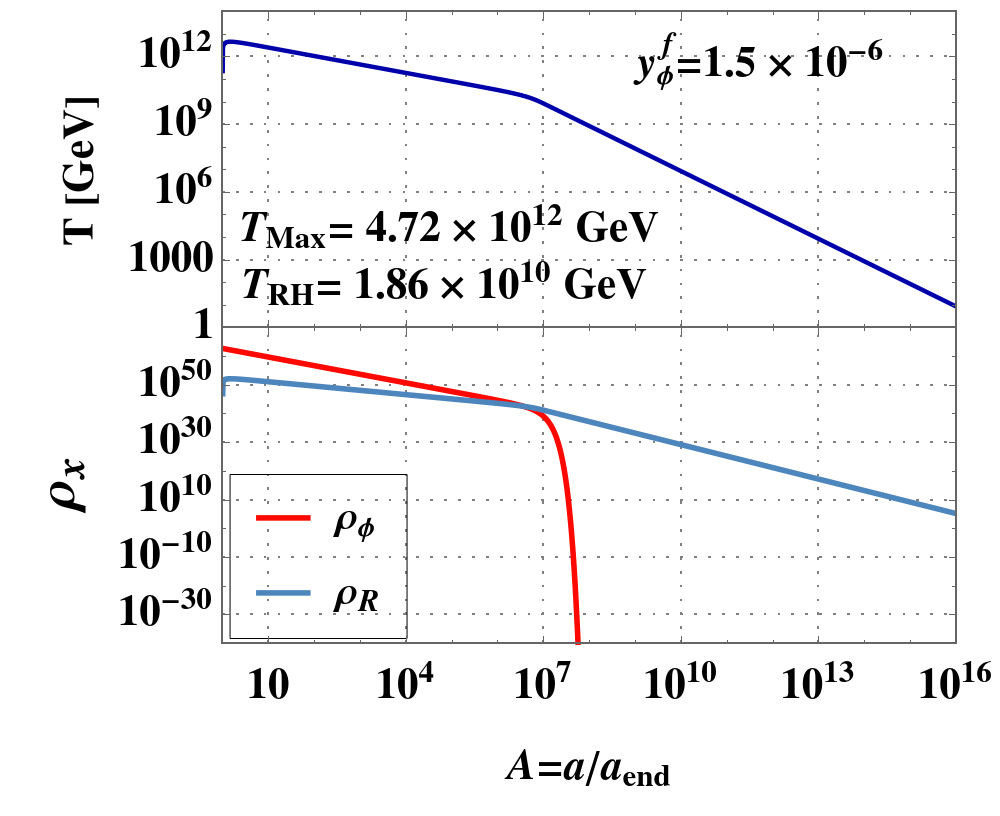}}
	\caption{Top (bottom) panel indicates evolutions of $T$ (various energy densities) as function of the relative scale factor.}
	\label{fig:2}
\end{figure}

The energy densities of the inflaton field ($\rho_{\phi}$) and radiation ($\rho_R$) satisfy the evolution equations\footnote{ Such scale factor dependance is explained in the Supplemental Material.}  in terms of $a$:
\begin{align}
	& \frac{d(\rho_\phi a^3)}{da}=- \frac{\Gamma_\phi}{\mathcal{H}}\rho_\phi a^2;~~{\rm{and}}~~ \frac{d\left(\rho_R a^4\right)}{da}= \frac{a^3}{\mathcal{H}} \Gamma_\phi \rho_\phi, 
	\label{rho-eq} 
\end{align}
where $\mathcal{H}^2 = ({\dot a}/a)^2 = (\rho_{\phi}+ \rho_{R})/3M^2_P$. At this moment, it is pertinent to define the temperature $T(a)$ connected to the $\rho_R$ by the relation
\begin{equation}
	T(a) = \left[ \frac{30}{g_*(a)\pi^2}\right]^{1/4} \rho^{1/4}_R(a).
\end{equation}
$T$ can therefore be estimated as a solution to the coupled Eqs. ~\eqref{rho-eq} whose variation with the relative scale factor ($a_{end}$ is the scale factor at the end of inflation, considered as a mere reference point) is presented in the top panel of Fig. {\ref{fig:2}} with $y^f_{\phi} = 1.5 \times 10^{-6}$ (with $g_f = 30$ kept throughout). Note that during the onset of reheating, $\mathcal{H}$ has the sole contribution from $\rho_{\phi}$. However, as soon as the light decay products of $\phi$ thermalizes, $\rho_R$ gets growing sharply and correspondingly a maximum temperature $T_{\text{Max}}$ is attained. After this, $\rho_R$ decreases slowly (as $\phi$ being decayed,  $\rho_{\phi}$ diminishes) and finally a stage is reached where $\rho_R$ eventually becomes equal to $\rho_{\phi}$ and starts dominating thereafter. This boundary is marked as the reheating temperature $T_{\text{RH}}$. 		

We now infer that provided the reheating temperature $T_{\rm{RH}}$ is smaller than the standard charged lepton Yukawa ET, $T^*_{0(\alpha)}$, for a specific flavor $\alpha$, the epoch in which this equilibration (by equating $\langle \Gamma_{\alpha} \rangle$ with $\mathcal{H}$) happens may actually experience the era between $T_{\text{Max}}$ and $T_{\text{RH}}$. The relatively faster expansion of the Universe in this era  will obviously affect the $r.h.s$ of Eq.~\eqref{eq-temp} so as to replace it with ${\mathcal{H}}$ which now depends on both $\rho_{\phi}$ and $\rho_{R}$. Using the $\rho_{\phi}$ and $\rho_R$ information obtained from Fig. {\ref{fig:2}} as input for ${\mathcal{H}}$, the modified $\langle \Gamma_{\alpha} \rangle/{\mathcal{H}}$ is plotted against $T$ in Fig. \ref{gamma-by-H} indicated by the dotted lines. As expected, prior to entering into the RD  phase, it depicts a non-standard behavior (slope is different within $T_{\text{Max}}$ and $T_{\text{RH}}$) as shown in Fig. \ref{gamma-by-H}, separately for three flavors of charged leptons. This is a new observation in our work, hitherto unexplored in the literature.

We note that $Y_{\tau}$ comes to equilibrium around $T^*_{\tau} \sim 5 \times 10^{10}$ GeV (compared to $T^*_{0({\tau})} \simeq 5 \times 10^{11}$ GeV in standard scenario) while $\mu_R$ and $e_R$ attain equilibriums at their standard temperatures $T^*_{0({\mu})}$ and $T^*_{0(e)}$ respectively in this case. This is because with the specific choice of $y_{\phi}^f = 1.5 \times 10^{-6}$, $T_{\text{RH}}$ is found to be $1.86 \times 10^{10}$ GeV which falls below $T^*_{0({\tau})}$. Hence the non-standard effect on $Y_{\mu}$ and $Y_{e}$ interactions do not appear in this particular example. For further smaller $y_{\phi}^f$, $T_{\text{RH}}$ may fall below $T^*_{0({\mu})}$ and correspondingly we would get a modified $Y_{\mu}$ ET. On the other hand, for sufficiently large $y_{\phi}^f$, the temperature regime between $T_{\text{Max}}$ and $T_{\text{RH}}$ is getting shortened and hence such departure in evaluating the charged lepton ET, even if it is present, becomes less prominent. In the Supplemental Material, we elaborate on it exploring a range of $y^f_{\phi}$.

Equipped with this basic understanding behind such a non-standard situation in case of prolonged reheating, we now turn our attention to the realm of flavor leptogenesis. The lepton asymmetry being generated from the out-of-equilibrium decay of the lightest RHN $N_1$, its mass $M_1$ is at the center of interest. This is because, in standard thermal leptogenesis, $N_1$ starts to decay at a temperature $T \sim M_1$ when the Universe is RD  and $N_1$ is present in equilibrium within the thermal bath. Contrary to this, in case of non-instantaneous reheating under study, in general there can be three possibilities: (a) $M_1 < T_{\text{RH}}$, (b) $T_{\text{RH}} < M_1 < T_{\text{Max}}$ and (c) $M_1 > T_{\text{Max}}$. 
While case (a) is a scenario close to standard thermal leptogenesis (and we refrain from discussing it here), case (c) stands for a situation where $N_1$ can not be thermally produced, discussed separately in another work \cite{Datta:2023pav}. Here we focus on the case (b), where thermal production of $N_1$ as well as its decay seem to be a possibility within this reheating epoch. 

To start, we need to introduce the energy density associated to $N_1$ ($\rho_{N_1}$) in the discussion as $T_{\text{Max}}$ being greater than $M_1$, $N_1$ can be produced from the inverse decay\footnote{For simplicity, we do not consider production via scattering.}.  The modified Hubble parameter is given by 
\begin{equation}
	\mathcal{H}^2 = \frac{\rho_{\phi}+ \rho_{R} +\rho_{N_1}}{3M^2_P}.
	\label{Hubble}
\end{equation}
Alongside, Eq.~\eqref{rho-eq} for $\rho_{R}$ would now have an additional term $\frac{a^3}{H} \langle\Gamma_{N_1}\rangle (\rho_{N_1}-\rho_{N_1}^{\text{eq}})$ on the $r.h.s.$, the presence of which is to represent the dilution of $\rho_R$ due to production of $N_1$ from it (via inverse decays). The decay rate of $N_1$ is given by $\langle \Gamma_{N_1}\rangle$ \cite{Buchmuller:2004nz} and  $\rho_{N_1}$ satisfies the other equation 
\begin{equation}
	\frac{d (\rho_{N_1} a^3)}{da}= -\frac{\langle\Gamma_{N_1}\rangle a^2}{\mathcal{H}}(\rho_{N_1}-\rho_{N_1}^{\text{eq}}).
\end{equation}

\begin{figure}[h]
	\centering
	{\includegraphics[width=1\linewidth]{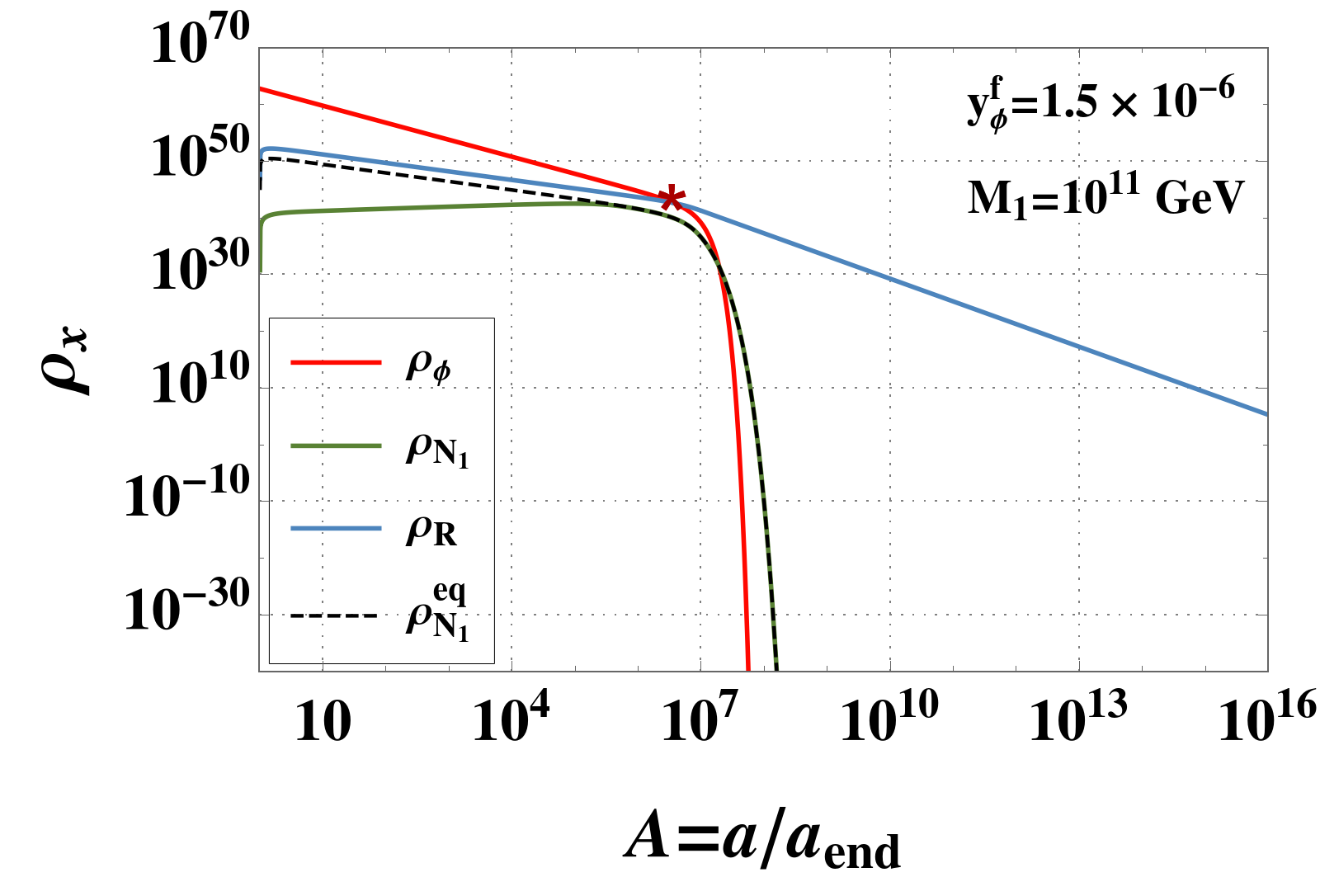}}
	\caption{Evolution of various components of total energy density are displayed. The $*$ indicates the $T_{\text{RH}}$.}
	\label{fig:3}
\end{figure}

Regarding initial conditions in order to solve the coupled equations for $\rho_{\phi}, \rho_R$ and $\rho_{N_1}$, we employ $\rho_{N_1} = \rho_R = 0$ whereas energy density at the end of inflation is taken to be $\rho_{\phi_{end}} = 3 V(\phi_{end})/2$. The value of $\phi_{end}$ for the class of models under consideration is $0.78 M_P$ \cite{Garcia:2020eof}. 
All the three components of the energy densities are now plotted against the relative scale factor in Fig. \ref{fig:3}. For this plot, we keep the parameter $y_{\phi}^f = 1.5 \times 10^{-6}$ same as in Fig. \ref{fig:2} and consider a specific $M_1 = 10^{11}$ GeV, lies in between $T_{\rm{Max}}$ and $T_{\rm{RH}}$ {\'a ~ la} case (b). We consider $M_{2(3)}$ as 100 times heavier than $M_{1(2)}$ for discussion and hence thermal generation of $N_{2(3)}$ is not possible.

The neutrino Yukawa $Y_{\nu}$ appearing in $\langle \Gamma_{N_1} \rangle$ is evaluated with the help of Casas-Ibarra formalism \cite{Casas:2001sr}, 
\begin{align}
	Y_{\nu}=-i \frac{\sqrt{2}}{v} U D_{\sqrt{m}} \mathbf{R} D_{\sqrt{M}},
\end{align}
where $U$ is the PMNS \cite{Esfahani:2017dmu,Esteban:2020cvm,Zyla:2020zbs} mixing matrix diagonalizing $m_{\nu} = - Y_{\nu} M^{-1}_R Y^T_{\nu} v^2/2$ ($v =$ 246 GeV is the EW vev), $D_m (D_M) $ is the diagonal active neutrino (RHN) mass matrix and $\rm{\bf{R}}$ is a complex orthogonal matrix of the form as in \cite{Antusch:2011nz} parametrized by complex mixing angle $\theta_R$. The values of $\rm{Re}[\theta_R], \rm{Im}[\theta_R]$ are so chosen as to produce correct baryon asymmetry via leptogenesis and also to keep $Y_{\nu}$ entries perturbative. We incorporate the best fit values of mixing angles and mass-squared differences \cite{Esteban:2020cvm} to define $U$ and light neutrino mass eigenvalues (with $m_1 = 0$). 

As can be seen from Fig. \ref{fig:3}, $N_1$ gradually increases, however, always remain subdominant compared to $\rho_{R}$. This is because it also decays out of equilibrium within the same period. As a result, the reheating temperature $T_{\text{RH}}$ is essentially unchanged even in presence of this new component, $\rho_{N_1}$. From Fig. \ref{gamma-by-H}, we recall that $Y_{\tau}$ enters equilibrium below $T^*_{\tau} = 5 \times 10^{10}$ GeV in this modified scenario. Therefore, when $N_1$ starts to decay, $i.e.$ around $T \lesssim M_1 (= 10^{11}$ GeV), none of the charged lepton Yukawa couplings reaches equilibrium. Hence, the flavor coherence is maintained and we encounter an unflavored regime of leptogenesis. Note that for the same $M_1$ value, in standard thermal leptogenesis happening in a RD era with $T^*_{0(\tau)} = 5 \times 10^{11}$ GeV being larger than $M_1$, $Y_\tau$-interaction would already be in equilibrium breaking flavor coherence. In that case, contrary to our realization, lepton asymmetries would be produced in two orthogonal directions: $\tau$ and $\kappa$, $\kappa$ forming the subspace of a coherent superposition of $e$ and $\mu$ lepton flavors \cite{Barbieri:1999ma,Nardi:2005hs,Abada:2006fw,Nardi:2006fx,Blanchet:2006be} leading to flavor leptogenesis. Such a shift in flavor regime (from two flavor to unflavored) of leptogenesis is an important outcome of our study.

Realizing that the case with $M_1 = 10^{11}$ GeV (and $y_{\phi}^f =1.5 \times 10^{-6}$) corresponds to unflavored leptogenesis scenario during this extended reheating period, we proceed to evaluate the $B-L$ asymmetry using the following Boltzmann equation {\color{blue}\cite{Buchmuller:2004nz}}, 
\begin{align}
	\frac{d (n_{\Delta} a^3)}{da}=-\frac{\langle\Gamma_{N_1}\rangle a^2}{\mathcal{H}}\left[\frac{\varepsilon_\ell}{M_1}(\rho_{N_1}-\rho_{N_1}^{\rm{eq}})+\frac{n_{N_1}^{\rm{eq}}}{2 n_\ell^{\rm{eq}}}n_{\Delta}\right],
\end{align}
with $n_{\Delta} = n_{B-L}$. However, in a more general case where a shift of regimes of thermal leptogenesis still leads to a flavored one, the corresponding equation would be
\begin{widetext}
	\begin{align} 
		\frac{d(n_{\Delta_i}a^3)}{da}=-\frac{\langle\Gamma_{N_1}\rangle a^2}{\mathcal{H}}\left[\frac{\varepsilon_{\ell_i}}{M_1}(\rho_{N_1}-\rho_{N_1}^{\text{eq}})
		+\frac{1}{2}K^0_i\sum_j(C^\ell_{ij}+C^H_{j})\frac{n_{N_1}^{\rm{eq}}}{n_\ell^{\rm{eq}}}n_{\Delta_j}\right].
	\end{align}
\end{widetext}
Here $K^0_{i}= (Y_{\nu}^*)_{\alpha 1} (Y_{\nu})_{\alpha 1}/{(Y_{\nu}^{\dagger} Y_{\nu})_{11}}$ is flavor projector \cite{Nardi:2006fx} and $C^{\ell}, C^H$ matrices 
connect the asymmetries in lepton and Higgs to asymmetries in $\Delta_{i} = B/3 - L_{i}$ (in terms of $n_{\Delta_{\ell_{\tau}}}$ and $n_{\Delta_{\ell_{\kappa}}}$ here)\cite{Nardi:2006fx}. The CP asymmetry $\varepsilon_{\ell_{\alpha}}$ (for unflavored case, $\varepsilon_{\ell} = \sum_{\alpha} \varepsilon_{\ell_{\alpha}}$) involved is obtained from the decay of $N_1$ to a specific flavor $\ell_{\alpha}$ and estimated using standard expression~\cite{Covi:1996wh,Nardi:2006fx}. The final baryon asymmetry $Y_{B}$ is related to $n_{\Delta_i}$ by~\cite{Harvey:1990qw}:
$Y_{B}= \frac{28}{79}\sum_\alpha n_{\Delta_\alpha}/s.$

\begin{figure}[h]
	\centering
	{\includegraphics[width=1\linewidth]{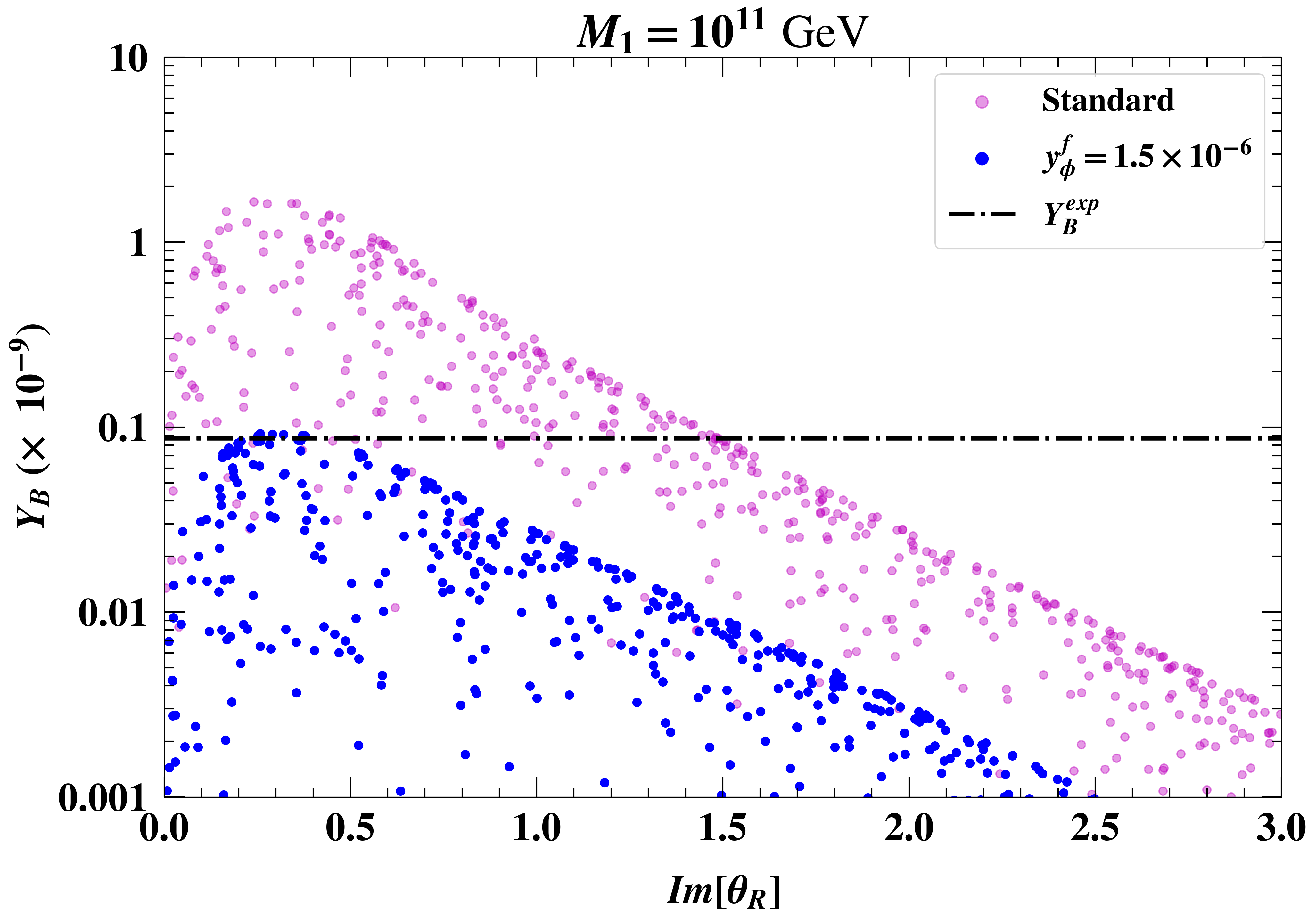}}
	\caption{$Y_B$ in case (a) non-instantaneous reheating (blue) and (b) standard RD phase (magenta): variation with ${\rm{Im}}[\theta_R]$ is displayed.}
	\label{fig:4}
\end{figure}
In Fig.~\ref{fig:4}, we compare the parameter space of $Y_{B}$ against ${\rm{Im}}[\theta_R]$, for the choices of $M_1$ and $y_{\phi}^f$ discussed above, obtained in usual thermal leptogenesis (fall in two flavored regimes, shown in magenta) with that in case of prolonged reheating (unflavored regime, shown in blue). Here we scan over the range of $\theta_R$ 
as: ${\rm{Re}}[\theta_R]$ (0 - $2 \pi$) and ${\rm{Im}}[\theta_R]$ (0-3). We observe a clear shift in the asymmetry  by almost an order of magnitude,signifying the effect of non-instantaneous reheating. Furthermore, in terms of the correct amount of baryon asymmetry, the situation becomes more restrictive as evident from the presence of a few number of blue points over magenta ones across the correct $Y_B$ line (horizontal black dashed). 
At this point, it is pertinent to inquire what happens if we choose a different set of $M_1, y_{\phi}^f$, within case (b). To answer it, we include a scan for $y_{\phi}^f-M_1$ in the Supplemental Material which helps make a clear distinction with the standard scenario in terms of finding the effect of the shift in the ET on thermal leptogenesis. Though we restrict ourselves in effective interaction between inflaton and SM fermions only, inflaton-boson couplings such as $\sigma \phi^2 H^{\dagger}H$ and $\mu \phi H^{\dagger}H$ may also be present. However, in order to keep decays of $\phi$ via such interactions in perturbative regime (to avoid rapid production of particles via preheating \cite{Kofman:1994rk}), the associated couplings should satisfy: $\sigma < 5.5 \times 10^{-12}, \mu < 5.2 \times 10^{-12}M_P$ \cite{Garcia:2020wiy}. It turns out that with such $\sigma$ and $\mu$ values, our analysis with $y^f_{\phi} \phi \bar{f}f$ alone would not be impacted by the presence of inflaton-boson coupling(s).

We have shown that the effect of a prolonged reheating period, starting from the maximum temperature of the Universe following inflation to the reheating, affects the equilibration of the charged lepton Yukawa interactions significantly. Due to the faster expansion of the Universe in this reheating period, the charged lepton Yukawa interactions enter equilibrium in a delayed fashion. This new observation depends on the effective coupling of the inflaton field with SM fields. With low reheating temperature ($i.e.$ with a smaller effective coupling of the inflaton field with SM fields), the era of reheating becomes longer and hence the ET(s) can be significantly smaller.
Effects of this observation should also be applicable to a wider SM interactions and open a possibility for further investigation in different directions. As one such application, we 
study here its effect on leptogenesis where the lightest RHN belonging to the type-I seesaw is being created and decayed out of equilibrium within $T_{\text{Max}}$ and $T_{\text{RH}}$. As expected, we find that the delayed entry of the charged lepton Yukawas in equilibrium can shift the flavor regimes of leptogenesis. This is a new observation and warrants more detailed studies. Alongside, there could be several directions that may open up. For example, in a follow-up study of this work, a new regime of flavor leptogenesis is found \cite{Datta:2023pav} in the context of non-thermal leptogenesis during extended reheating. Finally, we expect that any low scale leptogenesis scenario with low reheating temperature would be greatly impacted by our finding and can also be tested in current and near future experiments.

\section*{Acknowledgment}			
	AD is supported by the National Research Foundation of Korea (NRF) grant funded by the Korean government (2022R1A5A1030700) and the Department of Physics, Kyungpook National University. 
	AD also acknowledges the grant CRG/2021/005080  from SERB, Govt. of India during his stay at the Department of Physics, Indian Institute of Technology Guwahati where part of this work was completed. 
	RR acknowledges financial support from the STFC Consolidated Grant ST/T000775/1. AS acknowledges the support from grants CRG/2021/005080 and MTR/2021/000774 from SERB, Govt. of India. 
			
\bibliography{ref}

\begin{thebibliography}{57}%
\makeatletter
\providecommand \@ifxundefined [1]{%
 \@ifx{#1\undefined}
}%
\providecommand \@ifnum [1]{%
 \ifnum #1\expandafter \@firstoftwo
 \else \expandafter \@secondoftwo
 \fi
}%
\providecommand \@ifx [1]{%
 \ifx #1\expandafter \@firstoftwo
 \else \expandafter \@secondoftwo
 \fi
}%
\providecommand \natexlab [1]{#1}%
\providecommand \enquote  [1]{``#1''}%
\providecommand \bibnamefont  [1]{#1}%
\providecommand \bibfnamefont [1]{#1}%
\providecommand \citenamefont [1]{#1}%
\providecommand \href@noop [0]{\@secondoftwo}%
\providecommand \href [0]{\begingroup \@sanitize@url \@href}%
\providecommand \@href[1]{\@@startlink{#1}\@@href}%
\providecommand \@@href[1]{\endgroup#1\@@endlink}%
\providecommand \@sanitize@url [0]{\catcode `\\12\catcode `\$12\catcode
  `\&12\catcode `\#12\catcode `\^12\catcode `\_12\catcode `\%12\relax}%
\providecommand \@@startlink[1]{}%
\providecommand \@@endlink[0]{}%
\providecommand \url  [0]{\begingroup\@sanitize@url \@url }%
\providecommand \@url [1]{\endgroup\@href {#1}{\urlprefix }}%
\providecommand \urlprefix  [0]{URL }%
\providecommand \Eprint [0]{\href }%
\providecommand \doibase [0]{https://doi.org/}%
\providecommand \selectlanguage [0]{\@gobble}%
\providecommand \bibinfo  [0]{\@secondoftwo}%
\providecommand \bibfield  [0]{\@secondoftwo}%
\providecommand \translation [1]{[#1]}%
\providecommand \BibitemOpen [0]{}%
\providecommand \bibitemStop [0]{}%
\providecommand \bibitemNoStop [0]{.\EOS\space}%
\providecommand \EOS [0]{\spacefactor3000\relax}%
\providecommand \BibitemShut  [1]{\csname bibitem#1\endcsname}%
\let\auto@bib@innerbib\@empty
\bibitem [{\citenamefont {Aghanim}\ \emph {et~al.}(2020)\citenamefont {Aghanim}
  \emph {et~al.}}]{Planck:2018vyg}%
  \BibitemOpen
  \bibfield  {author} {\bibinfo {author} {\bibfnamefont {N.}~\bibnamefont
  {Aghanim}} \emph {et~al.} (\bibinfo {collaboration} {Planck}),\ }\bibfield
  {title} {\bibinfo {title} {{Planck 2018 results. VI. Cosmological
  parameters}},\ }\href {https://doi.org/10.1051/0004-6361/201833910}
  {\bibfield  {journal} {\bibinfo  {journal} {Astron. Astrophys.}\ }\textbf
  {\bibinfo {volume} {641}},\ \bibinfo {pages} {A6} (\bibinfo {year} {2020})},\
  \bibinfo {note} {[Erratum: Astron.Astrophys. 652, C4 (2021)]},\ \Eprint
  {https://arxiv.org/abs/1807.06209} {arXiv:1807.06209 [astro-ph.CO]}
  \BibitemShut {NoStop}%
\bibitem [{\citenamefont {Fukugita}\ and\ \citenamefont
  {Yanagida}(1986)}]{Fukugita:1986hr}%
  \BibitemOpen
  \bibfield  {author} {\bibinfo {author} {\bibfnamefont {M.}~\bibnamefont
  {Fukugita}}\ and\ \bibinfo {author} {\bibfnamefont {T.}~\bibnamefont
  {Yanagida}},\ }\bibfield  {title} {\bibinfo {title} {{Baryogenesis Without
  Grand Unification}},\ }\href {https://doi.org/10.1016/0370-2693(86)91126-3}
  {\bibfield  {journal} {\bibinfo  {journal} {Phys. Lett. B}\ }\textbf
  {\bibinfo {volume} {174}},\ \bibinfo {pages} {45} (\bibinfo {year}
  {1986})}\BibitemShut {NoStop}%
\bibitem [{\citenamefont {Luty}(1992)}]{Luty:1992un}%
  \BibitemOpen
  \bibfield  {author} {\bibinfo {author} {\bibfnamefont {M.~A.}\ \bibnamefont
  {Luty}},\ }\bibfield  {title} {\bibinfo {title} {{Baryogenesis via
  leptogenesis}},\ }\href {https://doi.org/10.1103/PhysRevD.45.455} {\bibfield
  {journal} {\bibinfo  {journal} {Phys. Rev. D}\ }\textbf {\bibinfo {volume}
  {45}},\ \bibinfo {pages} {455} (\bibinfo {year} {1992})}\BibitemShut
  {NoStop}%
\bibitem [{\citenamefont {Pilaftsis}(1997)}]{Pilaftsis:1997jf}%
  \BibitemOpen
  \bibfield  {author} {\bibinfo {author} {\bibfnamefont {A.}~\bibnamefont
  {Pilaftsis}},\ }\bibfield  {title} {\bibinfo {title} {{CP violation and
  baryogenesis due to heavy Majorana neutrinos}},\ }\href
  {https://doi.org/10.1103/PhysRevD.56.5431} {\bibfield  {journal} {\bibinfo
  {journal} {Phys. Rev. D}\ }\textbf {\bibinfo {volume} {56}},\ \bibinfo
  {pages} {5431} (\bibinfo {year} {1997})},\ \Eprint
  {https://arxiv.org/abs/hep-ph/9707235} {arXiv:hep-ph/9707235} \BibitemShut
  {NoStop}%
\bibitem [{\citenamefont {Minkowski}(1977)}]{Minkowski:1977sc}%
  \BibitemOpen
  \bibfield  {author} {\bibinfo {author} {\bibfnamefont {P.}~\bibnamefont
  {Minkowski}},\ }\bibfield  {title} {\bibinfo {title} {{$\mu \to e\gamma$ at a
  Rate of One Out of $10^{9}$ Muon Decays?}},\ }\href
  {https://doi.org/10.1016/0370-2693(77)90435-X} {\bibfield  {journal}
  {\bibinfo  {journal} {Phys. Lett. B}\ }\textbf {\bibinfo {volume} {67}},\
  \bibinfo {pages} {421} (\bibinfo {year} {1977})}\BibitemShut {NoStop}%
\bibitem [{\citenamefont {Yanagida}(1979{\natexlab{a}})}]{Yanagida:1979as}%
  \BibitemOpen
  \bibfield  {author} {\bibinfo {author} {\bibfnamefont {T.}~\bibnamefont
  {Yanagida}},\ }\bibfield  {title} {\bibinfo {title} {{Horizontal gauge
  symmetry and masses of neutrinos}},\ }\href@noop {} {\bibfield  {journal}
  {\bibinfo  {journal} {Conf. Proc. C}\ }\textbf {\bibinfo {volume}
  {7902131}},\ \bibinfo {pages} {95} (\bibinfo {year}
  {1979}{\natexlab{a}})}\BibitemShut {NoStop}%
\bibitem [{\citenamefont {Yanagida}(1979{\natexlab{b}})}]{Yanagida:1979gs}%
  \BibitemOpen
  \bibfield  {author} {\bibinfo {author} {\bibfnamefont {T.}~\bibnamefont
  {Yanagida}},\ }\bibfield  {title} {\bibinfo {title} {{Horizontal Symmetry and
  Mass of the Top Quark}},\ }\href {https://doi.org/10.1103/PhysRevD.20.2986}
  {\bibfield  {journal} {\bibinfo  {journal} {Phys. Rev. D}\ }\textbf {\bibinfo
  {volume} {20}},\ \bibinfo {pages} {2986} (\bibinfo {year}
  {1979}{\natexlab{b}})}\BibitemShut {NoStop}%
\bibitem [{\citenamefont {Gell-Mann}\ \emph {et~al.}(1979)\citenamefont
  {Gell-Mann}, \citenamefont {Ramond},\ and\ \citenamefont
  {Slansky}}]{GellMann:1980vs}%
  \BibitemOpen
  \bibfield  {author} {\bibinfo {author} {\bibfnamefont {M.}~\bibnamefont
  {Gell-Mann}}, \bibinfo {author} {\bibfnamefont {P.}~\bibnamefont {Ramond}},\
  and\ \bibinfo {author} {\bibfnamefont {R.}~\bibnamefont {Slansky}},\
  }\bibfield  {title} {\bibinfo {title} {{Complex Spinors and Unified
  Theories}},\ }\href@noop {} {\bibfield  {journal} {\bibinfo  {journal} {Conf.
  Proc. C}\ }\textbf {\bibinfo {volume} {790927}},\ \bibinfo {pages} {315}
  (\bibinfo {year} {1979})},\ \Eprint {https://arxiv.org/abs/1306.4669}
  {arXiv:1306.4669 [hep-th]} \BibitemShut {NoStop}%
\bibitem [{\citenamefont {Mohapatra}\ and\ \citenamefont
  {Senjanovic}(1980)}]{Mohapatra:1979ia}%
  \BibitemOpen
  \bibfield  {author} {\bibinfo {author} {\bibfnamefont {R.~N.}\ \bibnamefont
  {Mohapatra}}\ and\ \bibinfo {author} {\bibfnamefont {G.}~\bibnamefont
  {Senjanovic}},\ }\bibfield  {title} {\bibinfo {title} {{Neutrino Mass and
  Spontaneous Parity Nonconservation}},\ }\href
  {https://doi.org/10.1103/PhysRevLett.44.912} {\bibfield  {journal} {\bibinfo
  {journal} {Phys. Rev. Lett.}\ }\textbf {\bibinfo {volume} {44}},\ \bibinfo
  {pages} {912} (\bibinfo {year} {1980})}\BibitemShut {NoStop}%
\bibitem [{\citenamefont {Schechter}\ and\ \citenamefont
  {Valle}(1980)}]{Schechter:1980gr}%
  \BibitemOpen
  \bibfield  {author} {\bibinfo {author} {\bibfnamefont {J.}~\bibnamefont
  {Schechter}}\ and\ \bibinfo {author} {\bibfnamefont {J.~W.~F.}\ \bibnamefont
  {Valle}},\ }\bibfield  {title} {\bibinfo {title} {{Neutrino Masses in SU(2) x
  U(1) Theories}},\ }\href {https://doi.org/10.1103/PhysRevD.22.2227}
  {\bibfield  {journal} {\bibinfo  {journal} {Phys. Rev. D}\ }\textbf {\bibinfo
  {volume} {22}},\ \bibinfo {pages} {2227} (\bibinfo {year}
  {1980})}\BibitemShut {NoStop}%
\bibitem [{\citenamefont {Schechter}\ and\ \citenamefont
  {Valle}(1982)}]{Schechter:1981cv}%
  \BibitemOpen
  \bibfield  {author} {\bibinfo {author} {\bibfnamefont {J.}~\bibnamefont
  {Schechter}}\ and\ \bibinfo {author} {\bibfnamefont {J.~W.~F.}\ \bibnamefont
  {Valle}},\ }\bibfield  {title} {\bibinfo {title} {{Neutrino Decay and
  Spontaneous Violation of Lepton Number}},\ }\href
  {https://doi.org/10.1103/PhysRevD.25.774} {\bibfield  {journal} {\bibinfo
  {journal} {Phys. Rev. D}\ }\textbf {\bibinfo {volume} {25}},\ \bibinfo
  {pages} {774} (\bibinfo {year} {1982})}\BibitemShut {NoStop}%
\bibitem [{\citenamefont {Datta}\ \emph {et~al.}(2021)\citenamefont {Datta},
  \citenamefont {Roshan},\ and\ \citenamefont {Sil}}]{Datta:2021elq}%
  \BibitemOpen
  \bibfield  {author} {\bibinfo {author} {\bibfnamefont {A.}~\bibnamefont
  {Datta}}, \bibinfo {author} {\bibfnamefont {R.}~\bibnamefont {Roshan}},\ and\
  \bibinfo {author} {\bibfnamefont {A.}~\bibnamefont {Sil}},\ }\bibfield
  {title} {\bibinfo {title} {{Imprint of the Seesaw Mechanism on Feebly
  Interacting Dark Matter and the Baryon Asymmetry}},\ }\href
  {https://doi.org/10.1103/PhysRevLett.127.231801} {\bibfield  {journal}
  {\bibinfo  {journal} {Phys. Rev. Lett.}\ }\textbf {\bibinfo {volume} {127}},\
  \bibinfo {pages} {231801} (\bibinfo {year} {2021})},\ \Eprint
  {https://arxiv.org/abs/2104.02030} {arXiv:2104.02030 [hep-ph]} \BibitemShut
  {NoStop}%
\bibitem [{\citenamefont {Kuzmin}\ \emph {et~al.}(1985)\citenamefont {Kuzmin},
  \citenamefont {Rubakov},\ and\ \citenamefont {Shaposhnikov}}]{Kuzmin:1985mm}%
  \BibitemOpen
  \bibfield  {author} {\bibinfo {author} {\bibfnamefont {V.~A.}\ \bibnamefont
  {Kuzmin}}, \bibinfo {author} {\bibfnamefont {V.~A.}\ \bibnamefont
  {Rubakov}},\ and\ \bibinfo {author} {\bibfnamefont {M.~E.}\ \bibnamefont
  {Shaposhnikov}},\ }\bibfield  {title} {\bibinfo {title} {{On the Anomalous
  Electroweak Baryon Number Nonconservation in the Early Universe}},\ }\href
  {https://doi.org/10.1016/0370-2693(85)91028-7} {\bibfield  {journal}
  {\bibinfo  {journal} {Phys. Lett. B}\ }\textbf {\bibinfo {volume} {155}},\
  \bibinfo {pages} {36} (\bibinfo {year} {1985})}\BibitemShut {NoStop}%
\bibitem [{\citenamefont {Bento}(2003)}]{Bento:2003jv}%
  \BibitemOpen
  \bibfield  {author} {\bibinfo {author} {\bibfnamefont {L.}~\bibnamefont
  {Bento}},\ }\bibfield  {title} {\bibinfo {title} {{Sphaleron relaxation
  temperatures}},\ }\href {https://doi.org/10.1088/1475-7516/2003/11/002}
  {\bibfield  {journal} {\bibinfo  {journal} {JCAP}\ }\textbf {\bibinfo
  {volume} {11}},\ \bibinfo {pages} {002}},\ \Eprint
  {https://arxiv.org/abs/hep-ph/0304263} {arXiv:hep-ph/0304263} \BibitemShut
  {NoStop}%
\bibitem [{\citenamefont {D'Onofrio}\ \emph {et~al.}(2014)\citenamefont
  {D'Onofrio}, \citenamefont {Rummukainen},\ and\ \citenamefont
  {Tranberg}}]{DOnofrio:2014rug}%
  \BibitemOpen
  \bibfield  {author} {\bibinfo {author} {\bibfnamefont {M.}~\bibnamefont
  {D'Onofrio}}, \bibinfo {author} {\bibfnamefont {K.}~\bibnamefont
  {Rummukainen}},\ and\ \bibinfo {author} {\bibfnamefont {A.}~\bibnamefont
  {Tranberg}},\ }\bibfield  {title} {\bibinfo {title} {{Sphaleron Rate in the
  Minimal Standard Model}},\ }\href
  {https://doi.org/10.1103/PhysRevLett.113.141602} {\bibfield  {journal}
  {\bibinfo  {journal} {Phys. Rev. Lett.}\ }\textbf {\bibinfo {volume} {113}},\
  \bibinfo {pages} {141602} (\bibinfo {year} {2014})},\ \Eprint
  {https://arxiv.org/abs/1404.3565} {arXiv:1404.3565 [hep-ph]} \BibitemShut
  {NoStop}%
\bibitem [{\citenamefont {Arnold}\ \emph {et~al.}(1997)\citenamefont {Arnold},
  \citenamefont {Son},\ and\ \citenamefont {Yaffe}}]{Arnold:1996dy}%
  \BibitemOpen
  \bibfield  {author} {\bibinfo {author} {\bibfnamefont {P.~B.}\ \bibnamefont
  {Arnold}}, \bibinfo {author} {\bibfnamefont {D.}~\bibnamefont {Son}},\ and\
  \bibinfo {author} {\bibfnamefont {L.~G.}\ \bibnamefont {Yaffe}},\ }\bibfield
  {title} {\bibinfo {title} {{The Hot baryon violation rate is O (alpha-w**5
  T**4)}},\ }\href {https://doi.org/10.1103/PhysRevD.55.6264} {\bibfield
  {journal} {\bibinfo  {journal} {Phys. Rev. D}\ }\textbf {\bibinfo {volume}
  {55}},\ \bibinfo {pages} {6264} (\bibinfo {year} {1997})},\ \Eprint
  {https://arxiv.org/abs/hep-ph/9609481} {arXiv:hep-ph/9609481} \BibitemShut
  {NoStop}%
\bibitem [{\citenamefont {Bodeker}(1998)}]{Bodeker:1998hm}%
  \BibitemOpen
  \bibfield  {author} {\bibinfo {author} {\bibfnamefont {D.}~\bibnamefont
  {Bodeker}},\ }\bibfield  {title} {\bibinfo {title} {{On the effective
  dynamics of soft nonAbelian gauge fields at finite temperature}},\ }\href
  {https://doi.org/10.1016/S0370-2693(98)00279-2} {\bibfield  {journal}
  {\bibinfo  {journal} {Phys. Lett. B}\ }\textbf {\bibinfo {volume} {426}},\
  \bibinfo {pages} {351} (\bibinfo {year} {1998})},\ \Eprint
  {https://arxiv.org/abs/hep-ph/9801430} {arXiv:hep-ph/9801430} \BibitemShut
  {NoStop}%
\bibitem [{\citenamefont {Arnold}\ \emph {et~al.}(1999)\citenamefont {Arnold},
  \citenamefont {Son},\ and\ \citenamefont {Yaffe}}]{Arnold:1998cy}%
  \BibitemOpen
  \bibfield  {author} {\bibinfo {author} {\bibfnamefont {P.~B.}\ \bibnamefont
  {Arnold}}, \bibinfo {author} {\bibfnamefont {D.~T.}\ \bibnamefont {Son}},\
  and\ \bibinfo {author} {\bibfnamefont {L.~G.}\ \bibnamefont {Yaffe}},\
  }\bibfield  {title} {\bibinfo {title} {{Effective dynamics of hot, soft
  nonAbelian gauge fields. Color conductivity and log(1/alpha) effects}},\
  }\href {https://doi.org/10.1103/PhysRevD.59.105020} {\bibfield  {journal}
  {\bibinfo  {journal} {Phys. Rev. D}\ }\textbf {\bibinfo {volume} {59}},\
  \bibinfo {pages} {105020} (\bibinfo {year} {1999})},\ \Eprint
  {https://arxiv.org/abs/hep-ph/9810216} {arXiv:hep-ph/9810216} \BibitemShut
  {NoStop}%
\bibitem [{\citenamefont {Kolb}\ and\ \citenamefont
  {Turner}(1990)}]{Kolb:1990vq}%
  \BibitemOpen
  \bibfield  {author} {\bibinfo {author} {\bibfnamefont {E.~W.}\ \bibnamefont
  {Kolb}}\ and\ \bibinfo {author} {\bibfnamefont {M.~S.}\ \bibnamefont
  {Turner}},\ }\href {https://doi.org/10.1201/9780429492860} {\emph {\bibinfo
  {title} {{The Early Universe}}}},\ Vol.~\bibinfo {volume} {69}\ (\bibinfo
  {year} {1990})\BibitemShut {NoStop}%
\bibitem [{\citenamefont {Nardi}\ \emph
  {et~al.}(2006{\natexlab{a}})\citenamefont {Nardi}, \citenamefont {Nir},
  \citenamefont {Racker},\ and\ \citenamefont {Roulet}}]{Nardi:2005hs}%
  \BibitemOpen
  \bibfield  {author} {\bibinfo {author} {\bibfnamefont {E.}~\bibnamefont
  {Nardi}}, \bibinfo {author} {\bibfnamefont {Y.}~\bibnamefont {Nir}}, \bibinfo
  {author} {\bibfnamefont {J.}~\bibnamefont {Racker}},\ and\ \bibinfo {author}
  {\bibfnamefont {E.}~\bibnamefont {Roulet}},\ }\bibfield  {title} {\bibinfo
  {title} {{On Higgs and sphaleron effects during the leptogenesis era}},\
  }\href {https://doi.org/10.1088/1126-6708/2006/01/068} {\bibfield  {journal}
  {\bibinfo  {journal} {JHEP}\ }\textbf {\bibinfo {volume} {01}},\ \bibinfo
  {pages} {068}},\ \Eprint {https://arxiv.org/abs/hep-ph/0512052}
  {arXiv:hep-ph/0512052} \BibitemShut {NoStop}%
\bibitem [{\citenamefont {Cline}\ \emph {et~al.}(1993)\citenamefont {Cline},
  \citenamefont {Kainulainen},\ and\ \citenamefont {Olive}}]{Cline:1993vv}%
  \BibitemOpen
  \bibfield  {author} {\bibinfo {author} {\bibfnamefont {J.~M.}\ \bibnamefont
  {Cline}}, \bibinfo {author} {\bibfnamefont {K.}~\bibnamefont {Kainulainen}},\
  and\ \bibinfo {author} {\bibfnamefont {K.~A.}\ \bibnamefont {Olive}},\
  }\bibfield  {title} {\bibinfo {title} {{On the erasure and regeneration of
  the primordial baryon asymmetry by sphalerons}},\ }\href
  {https://doi.org/10.1103/PhysRevLett.71.2372} {\bibfield  {journal} {\bibinfo
   {journal} {Phys. Rev. Lett.}\ }\textbf {\bibinfo {volume} {71}},\ \bibinfo
  {pages} {2372} (\bibinfo {year} {1993})},\ \Eprint
  {https://arxiv.org/abs/hep-ph/9304321} {arXiv:hep-ph/9304321} \BibitemShut
  {NoStop}%
\bibitem [{\citenamefont {Cline}\ \emph {et~al.}(1994)\citenamefont {Cline},
  \citenamefont {Kainulainen},\ and\ \citenamefont {Olive}}]{Cline:1993bd}%
  \BibitemOpen
  \bibfield  {author} {\bibinfo {author} {\bibfnamefont {J.~M.}\ \bibnamefont
  {Cline}}, \bibinfo {author} {\bibfnamefont {K.}~\bibnamefont {Kainulainen}},\
  and\ \bibinfo {author} {\bibfnamefont {K.~A.}\ \bibnamefont {Olive}},\
  }\bibfield  {title} {\bibinfo {title} {{Protecting the primordial baryon
  asymmetry from erasure by sphalerons}},\ }\href
  {https://doi.org/10.1103/PhysRevD.49.6394} {\bibfield  {journal} {\bibinfo
  {journal} {Phys. Rev. D}\ }\textbf {\bibinfo {volume} {49}},\ \bibinfo
  {pages} {6394} (\bibinfo {year} {1994})},\ \Eprint
  {https://arxiv.org/abs/hep-ph/9401208} {arXiv:hep-ph/9401208} \BibitemShut
  {NoStop}%
\bibitem [{\citenamefont {Weldon}(1982)}]{Weldon:1982bn}%
  \BibitemOpen
  \bibfield  {author} {\bibinfo {author} {\bibfnamefont {H.~A.}\ \bibnamefont
  {Weldon}},\ }\bibfield  {title} {\bibinfo {title} {{Effective Fermion Masses
  of Order gT in High Temperature Gauge Theories with Exact Chiral
  Invariance}},\ }\href {https://doi.org/10.1103/PhysRevD.26.2789} {\bibfield
  {journal} {\bibinfo  {journal} {Phys. Rev. D}\ }\textbf {\bibinfo {volume}
  {26}},\ \bibinfo {pages} {2789} (\bibinfo {year} {1982})}\BibitemShut
  {NoStop}%
\bibitem [{\citenamefont {Quiros}(1999)}]{Quiros:1999jp}%
  \BibitemOpen
  \bibfield  {author} {\bibinfo {author} {\bibfnamefont {M.}~\bibnamefont
  {Quiros}},\ }\bibfield  {title} {\bibinfo {title} {{Finite temperature field
  theory and phase transitions}},\ }in\ \href@noop {} { {\bibinfo
  {booktitle} {{ICTP Summer School in High-Energy Physics and Cosmology}}}}\
  (\bibinfo {year} {1999})\ pp.\ \bibinfo {pages} {187--259},\ \Eprint
  {https://arxiv.org/abs/hep-ph/9901312} {arXiv:hep-ph/9901312} \BibitemShut
  {NoStop}%
\bibitem [{\citenamefont {Senaha}(2020)}]{Senaha:2020mop}%
  \BibitemOpen
  \bibfield  {author} {\bibinfo {author} {\bibfnamefont {E.}~\bibnamefont
  {Senaha}},\ }\bibfield  {title} {\bibinfo {title} {{Symmetry Restoration and
  Breaking at Finite Temperature: An Introductory Review}},\ }\href
  {https://doi.org/10.3390/sym12050733} {\bibfield  {journal} {\bibinfo
  {journal} {Symmetry}\ }\textbf {\bibinfo {volume} {12}},\ \bibinfo {pages}
  {733} (\bibinfo {year} {2020})}\BibitemShut {NoStop}%
\bibitem [{\citenamefont {Garbrecht}\ \emph {et~al.}(2013)\citenamefont
  {Garbrecht}, \citenamefont {Glowna},\ and\ \citenamefont
  {Schwaller}}]{Garbrecht:2013bia}%
  \BibitemOpen
  \bibfield  {author} {\bibinfo {author} {\bibfnamefont {B.}~\bibnamefont
  {Garbrecht}}, \bibinfo {author} {\bibfnamefont {F.}~\bibnamefont {Glowna}},\
  and\ \bibinfo {author} {\bibfnamefont {P.}~\bibnamefont {Schwaller}},\
  }\bibfield  {title} {\bibinfo {title} {{Scattering Rates For Leptogenesis:
  Damping of Lepton Flavour Coherence and Production of Singlet Neutrinos}},\
  }\href {https://doi.org/10.1016/j.nuclphysb.2013.08.020} {\bibfield
  {journal} {\bibinfo  {journal} {Nucl. Phys. B}\ }\textbf {\bibinfo {volume}
  {877}},\ \bibinfo {pages} {1} (\bibinfo {year} {2013})},\ \Eprint
  {https://arxiv.org/abs/1303.5498} {arXiv:1303.5498 [hep-ph]} \BibitemShut
  {NoStop}%
\bibitem [{\citenamefont {B\"odeker}\ and\ \citenamefont
  {Schr\"oder}(2019)}]{Bodeker:2019ajh}%
  \BibitemOpen
  \bibfield  {author} {\bibinfo {author} {\bibfnamefont {D.}~\bibnamefont
  {B\"odeker}}\ and\ \bibinfo {author} {\bibfnamefont {D.}~\bibnamefont
  {Schr\"oder}},\ }\bibfield  {title} {\bibinfo {title} {{Equilibration of
  right-handed electrons}},\ }\href
  {https://doi.org/10.1088/1475-7516/2019/05/010} {\bibfield  {journal}
  {\bibinfo  {journal} {JCAP}\ }\textbf {\bibinfo {volume} {05}},\ \bibinfo
  {pages} {010}},\ \Eprint {https://arxiv.org/abs/1902.07220} {arXiv:1902.07220
  [hep-ph]} \BibitemShut {NoStop}%
\bibitem [{\citenamefont {Barbieri}\ \emph {et~al.}(2000)\citenamefont
  {Barbieri}, \citenamefont {Creminelli}, \citenamefont {Strumia},\ and\
  \citenamefont {Tetradis}}]{Barbieri:1999ma}%
  \BibitemOpen
  \bibfield  {author} {\bibinfo {author} {\bibfnamefont {R.}~\bibnamefont
  {Barbieri}}, \bibinfo {author} {\bibfnamefont {P.}~\bibnamefont
  {Creminelli}}, \bibinfo {author} {\bibfnamefont {A.}~\bibnamefont
  {Strumia}},\ and\ \bibinfo {author} {\bibfnamefont {N.}~\bibnamefont
  {Tetradis}},\ }\bibfield  {title} {\bibinfo {title} {{Baryogenesis through
  leptogenesis}},\ }\href {https://doi.org/10.1016/S0550-3213(00)00011-0}
  {\bibfield  {journal} {\bibinfo  {journal} {Nucl. Phys. B}\ }\textbf
  {\bibinfo {volume} {575}},\ \bibinfo {pages} {61} (\bibinfo {year} {2000})},\
  \Eprint {https://arxiv.org/abs/hep-ph/9911315} {arXiv:hep-ph/9911315}
  \BibitemShut {NoStop}%
\bibitem [{\citenamefont {Abada}\ \emph {et~al.}(2006)\citenamefont {Abada},
  \citenamefont {Davidson}, \citenamefont {Josse-Michaux}, \citenamefont
  {Losada},\ and\ \citenamefont {Riotto}}]{Abada:2006fw}%
  \BibitemOpen
  \bibfield  {author} {\bibinfo {author} {\bibfnamefont {A.}~\bibnamefont
  {Abada}}, \bibinfo {author} {\bibfnamefont {S.}~\bibnamefont {Davidson}},
  \bibinfo {author} {\bibfnamefont {F.-X.}\ \bibnamefont {Josse-Michaux}},
  \bibinfo {author} {\bibfnamefont {M.}~\bibnamefont {Losada}},\ and\ \bibinfo
  {author} {\bibfnamefont {A.}~\bibnamefont {Riotto}},\ }\bibfield  {title}
  {\bibinfo {title} {{Flavor issues in leptogenesis}},\ }\href
  {https://doi.org/10.1088/1475-7516/2006/04/004} {\bibfield  {journal}
  {\bibinfo  {journal} {JCAP}\ }\textbf {\bibinfo {volume} {04}},\ \bibinfo
  {pages} {004}},\ \Eprint {https://arxiv.org/abs/hep-ph/0601083}
  {arXiv:hep-ph/0601083} \BibitemShut {NoStop}%
\bibitem [{\citenamefont {Nardi}\ \emph
  {et~al.}(2006{\natexlab{b}})\citenamefont {Nardi}, \citenamefont {Nir},
  \citenamefont {Roulet},\ and\ \citenamefont {Racker}}]{Nardi:2006fx}%
  \BibitemOpen
  \bibfield  {author} {\bibinfo {author} {\bibfnamefont {E.}~\bibnamefont
  {Nardi}}, \bibinfo {author} {\bibfnamefont {Y.}~\bibnamefont {Nir}}, \bibinfo
  {author} {\bibfnamefont {E.}~\bibnamefont {Roulet}},\ and\ \bibinfo {author}
  {\bibfnamefont {J.}~\bibnamefont {Racker}},\ }\bibfield  {title} {\bibinfo
  {title} {{The Importance of flavor in leptogenesis}},\ }\href
  {https://doi.org/10.1088/1126-6708/2006/01/164} {\bibfield  {journal}
  {\bibinfo  {journal} {JHEP}\ }\textbf {\bibinfo {volume} {01}},\ \bibinfo
  {pages} {164}},\ \Eprint {https://arxiv.org/abs/hep-ph/0601084}
  {arXiv:hep-ph/0601084} \BibitemShut {NoStop}%
\bibitem [{\citenamefont {Blanchet}\ and\ \citenamefont
  {Di~Bari}(2007)}]{Blanchet:2006be}%
  \BibitemOpen
  \bibfield  {author} {\bibinfo {author} {\bibfnamefont {S.}~\bibnamefont
  {Blanchet}}\ and\ \bibinfo {author} {\bibfnamefont {P.}~\bibnamefont
  {Di~Bari}},\ }\bibfield  {title} {\bibinfo {title} {{Flavor effects on
  leptogenesis predictions}},\ }\href
  {https://doi.org/10.1088/1475-7516/2007/03/018} {\bibfield  {journal}
  {\bibinfo  {journal} {JCAP}\ }\textbf {\bibinfo {volume} {03}},\ \bibinfo
  {pages} {018}},\ \Eprint {https://arxiv.org/abs/hep-ph/0607330}
  {arXiv:hep-ph/0607330} \BibitemShut {NoStop}%
\bibitem [{\citenamefont {Dev}\ \emph {et~al.}(2018)\citenamefont {Dev},
  \citenamefont {Di~Bari}, \citenamefont {Garbrecht}, \citenamefont {Lavignac},
  \citenamefont {Millington},\ and\ \citenamefont {Teresi}}]{Dev:2017trv}%
  \BibitemOpen
  \bibfield  {author} {\bibinfo {author} {\bibfnamefont {P.~S.~B.}\
  \bibnamefont {Dev}}, \bibinfo {author} {\bibfnamefont {P.}~\bibnamefont
  {Di~Bari}}, \bibinfo {author} {\bibfnamefont {B.}~\bibnamefont {Garbrecht}},
  \bibinfo {author} {\bibfnamefont {S.}~\bibnamefont {Lavignac}}, \bibinfo
  {author} {\bibfnamefont {P.}~\bibnamefont {Millington}},\ and\ \bibinfo
  {author} {\bibfnamefont {D.}~\bibnamefont {Teresi}},\ }\bibfield  {title}
  {\bibinfo {title} {{Flavor effects in leptogenesis}},\ }\href
  {https://doi.org/10.1142/S0217751X18420010} {\bibfield  {journal} {\bibinfo
  {journal} {Int. J. Mod. Phys. A}\ }\textbf {\bibinfo {volume} {33}},\
  \bibinfo {pages} {1842001} (\bibinfo {year} {2018})},\ \Eprint
  {https://arxiv.org/abs/1711.02861} {arXiv:1711.02861 [hep-ph]} \BibitemShut
  {NoStop}%
\bibitem [{\citenamefont {Giudice}\ \emph {et~al.}(2001)\citenamefont
  {Giudice}, \citenamefont {Kolb},\ and\ \citenamefont
  {Riotto}}]{Giudice:2000ex}%
  \BibitemOpen
  \bibfield  {author} {\bibinfo {author} {\bibfnamefont {G.~F.}\ \bibnamefont
  {Giudice}}, \bibinfo {author} {\bibfnamefont {E.~W.}\ \bibnamefont {Kolb}},\
  and\ \bibinfo {author} {\bibfnamefont {A.}~\bibnamefont {Riotto}},\
  }\bibfield  {title} {\bibinfo {title} {{Largest temperature of the radiation
  era and its cosmological implications}},\ }\href
  {https://doi.org/10.1103/PhysRevD.64.023508} {\bibfield  {journal} {\bibinfo
  {journal} {Phys. Rev. D}\ }\textbf {\bibinfo {volume} {64}},\ \bibinfo
  {pages} {023508} (\bibinfo {year} {2001})},\ \Eprint
  {https://arxiv.org/abs/hep-ph/0005123} {arXiv:hep-ph/0005123} \BibitemShut
  {NoStop}%
\bibitem [{\citenamefont {Garcia}\ \emph {et~al.}(2020)\citenamefont {Garcia},
  \citenamefont {Kaneta}, \citenamefont {Mambrini},\ and\ \citenamefont
  {Olive}}]{Garcia:2020eof}%
  \BibitemOpen
  \bibfield  {author} {\bibinfo {author} {\bibfnamefont {M.~A.~G.}\
  \bibnamefont {Garcia}}, \bibinfo {author} {\bibfnamefont {K.}~\bibnamefont
  {Kaneta}}, \bibinfo {author} {\bibfnamefont {Y.}~\bibnamefont {Mambrini}},\
  and\ \bibinfo {author} {\bibfnamefont {K.~A.}\ \bibnamefont {Olive}},\
  }\bibfield  {title} {\bibinfo {title} {{Reheating and Post-inflationary
  Production of Dark Matter}},\ }\href
  {https://doi.org/10.1103/PhysRevD.101.123507} {\bibfield  {journal} {\bibinfo
   {journal} {Phys. Rev. D}\ }\textbf {\bibinfo {volume} {101}},\ \bibinfo
  {pages} {123507} (\bibinfo {year} {2020})},\ \Eprint
  {https://arxiv.org/abs/2004.08404} {arXiv:2004.08404 [hep-ph]} \BibitemShut
  {NoStop}%
\bibitem [{\citenamefont {Antusch}\ and\ \citenamefont
  {Teixeira}(2007)}]{Antusch:2006gy}%
  \BibitemOpen
  \bibfield  {author} {\bibinfo {author} {\bibfnamefont {S.}~\bibnamefont
  {Antusch}}\ and\ \bibinfo {author} {\bibfnamefont {A.~M.}\ \bibnamefont
  {Teixeira}},\ }\bibfield  {title} {\bibinfo {title} {{Towards constraints on
  the SUSY seesaw from flavour-dependent leptogenesis}},\ }\href
  {https://doi.org/10.1088/1475-7516/2007/02/024} {\bibfield  {journal}
  {\bibinfo  {journal} {JCAP}\ }\textbf {\bibinfo {volume} {02}},\ \bibinfo
  {pages} {024}},\ \Eprint {https://arxiv.org/abs/hep-ph/0611232}
  {arXiv:hep-ph/0611232} \BibitemShut {NoStop}%
\bibitem [{\citenamefont {Antusch}(2007)}]{Antusch:2007km}%
  \BibitemOpen
  \bibfield  {author} {\bibinfo {author} {\bibfnamefont {S.}~\bibnamefont
  {Antusch}},\ }\bibfield  {title} {\bibinfo {title} {{Flavour-dependent type
  II leptogenesis}},\ }\href {https://doi.org/10.1103/PhysRevD.76.023512}
  {\bibfield  {journal} {\bibinfo  {journal} {Phys. Rev. D}\ }\textbf {\bibinfo
  {volume} {76}},\ \bibinfo {pages} {023512} (\bibinfo {year} {2007})},\
  \Eprint {https://arxiv.org/abs/0704.1591} {arXiv:0704.1591 [hep-ph]}
  \BibitemShut {NoStop}%
\bibitem [{\citenamefont {Mahanta}\ and\ \citenamefont
  {Borah}(2020)}]{Mahanta:2019sfo}%
  \BibitemOpen
  \bibfield  {author} {\bibinfo {author} {\bibfnamefont {D.}~\bibnamefont
  {Mahanta}}\ and\ \bibinfo {author} {\bibfnamefont {D.}~\bibnamefont
  {Borah}},\ }\bibfield  {title} {\bibinfo {title} {{TeV Scale Leptogenesis
  with Dark Matter in Non-standard Cosmology}},\ }\href
  {https://doi.org/10.1088/1475-7516/2020/04/032} {\bibfield  {journal}
  {\bibinfo  {journal} {JCAP}\ }\textbf {\bibinfo {volume} {04}}\bibfield
  {number} {\bibinfo  {number} { (04)},\ \bibinfo {pages} {032}},\ }\Eprint
  {https://arxiv.org/abs/1912.09726} {arXiv:1912.09726 [hep-ph]} \BibitemShut
  {NoStop}%
\bibitem [{\citenamefont {Perez-Gonzalez}\ and\ \citenamefont
  {Turner}(2021)}]{Perez-Gonzalez:2020vnz}%
  \BibitemOpen
  \bibfield  {author} {\bibinfo {author} {\bibfnamefont {Y.~F.}\ \bibnamefont
  {Perez-Gonzalez}}\ and\ \bibinfo {author} {\bibfnamefont {J.}~\bibnamefont
  {Turner}},\ }\bibfield  {title} {\bibinfo {title} {{Assessing the tension
  between a black hole dominated early universe and leptogenesis}},\ }\href
  {https://doi.org/10.1103/PhysRevD.104.103021} {\bibfield  {journal} {\bibinfo
   {journal} {Phys. Rev. D}\ }\textbf {\bibinfo {volume} {104}},\ \bibinfo
  {pages} {103021} (\bibinfo {year} {2021})},\ \Eprint
  {https://arxiv.org/abs/2010.03565} {arXiv:2010.03565 [hep-ph]} \BibitemShut
  {NoStop}%
\bibitem [{\citenamefont {Kallosh}\ and\ \citenamefont
  {Linde}(2013)}]{Kallosh:2013hoa}%
  \BibitemOpen
  \bibfield  {author} {\bibinfo {author} {\bibfnamefont {R.}~\bibnamefont
  {Kallosh}}\ and\ \bibinfo {author} {\bibfnamefont {A.}~\bibnamefont
  {Linde}},\ }\bibfield  {title} {\bibinfo {title} {{Universality Class in
  Conformal Inflation}},\ }\href
  {https://doi.org/10.1088/1475-7516/2013/07/002} {\bibfield  {journal}
  {\bibinfo  {journal} {JCAP}\ }\textbf {\bibinfo {volume} {07}},\ \bibinfo
  {pages} {002}},\ \Eprint {https://arxiv.org/abs/1306.5220} {arXiv:1306.5220
  [hep-th]} \BibitemShut {NoStop}%
\bibitem [{\citenamefont {Starobinsky}(1980)}]{Starobinsky:1980te}%
  \BibitemOpen
  \bibfield  {author} {\bibinfo {author} {\bibfnamefont {A.~A.}\ \bibnamefont
  {Starobinsky}},\ }\bibfield  {title} {\bibinfo {title} {{A New Type of
  Isotropic Cosmological Models Without Singularity}},\ }\href
  {https://doi.org/10.1016/0370-2693(80)90670-X} {\bibfield  {journal}
  {\bibinfo  {journal} {Phys. Lett. B}\ }\textbf {\bibinfo {volume} {91}},\
  \bibinfo {pages} {99} (\bibinfo {year} {1980})}\BibitemShut {NoStop}%
\bibitem [{\citenamefont {Ellis}\ \emph {et~al.}(2013)\citenamefont {Ellis},
  \citenamefont {Nanopoulos},\ and\ \citenamefont {Olive}}]{Ellis:2013xoa}%
  \BibitemOpen
  \bibfield  {author} {\bibinfo {author} {\bibfnamefont {J.}~\bibnamefont
  {Ellis}}, \bibinfo {author} {\bibfnamefont {D.~V.}\ \bibnamefont
  {Nanopoulos}},\ and\ \bibinfo {author} {\bibfnamefont {K.~A.}\ \bibnamefont
  {Olive}},\ }\bibfield  {title} {\bibinfo {title} {{No-Scale Supergravity
  Realization of the Starobinsky Model of Inflation}},\ }\href
  {https://doi.org/10.1103/PhysRevLett.111.111301} {\bibfield  {journal}
  {\bibinfo  {journal} {Phys. Rev. Lett.}\ }\textbf {\bibinfo {volume} {111}},\
  \bibinfo {pages} {111301} (\bibinfo {year} {2013})},\ \bibinfo {note}
  {[Erratum: Phys.Rev.Lett. 111, 129902 (2013)]},\ \Eprint
  {https://arxiv.org/abs/1305.1247} {arXiv:1305.1247 [hep-th]} \BibitemShut
  {NoStop}%
\bibitem [{\citenamefont {Khalil}\ \emph {et~al.}(2019)\citenamefont {Khalil},
  \citenamefont {Moursy}, \citenamefont {Saha},\ and\ \citenamefont
  {Sil}}]{Khalil:2018iip}%
  \BibitemOpen
  \bibfield  {author} {\bibinfo {author} {\bibfnamefont {S.}~\bibnamefont
  {Khalil}}, \bibinfo {author} {\bibfnamefont {A.}~\bibnamefont {Moursy}},
  \bibinfo {author} {\bibfnamefont {A.~K.}\ \bibnamefont {Saha}},\ and\
  \bibinfo {author} {\bibfnamefont {A.}~\bibnamefont {Sil}},\ }\bibfield
  {title} {\bibinfo {title} {{U(1)R inspired inflation model in no-scale
  supergravity}},\ }\href {https://doi.org/10.1103/PhysRevD.99.095022}
  {\bibfield  {journal} {\bibinfo  {journal} {Phys. Rev. D}\ }\textbf {\bibinfo
  {volume} {99}},\ \bibinfo {pages} {095022} (\bibinfo {year} {2019})},\
  \Eprint {https://arxiv.org/abs/1810.06408} {arXiv:1810.06408 [hep-ph]}
  \BibitemShut {NoStop}%
\bibitem [{\citenamefont {Ellis}\ \emph {et~al.}(2015)\citenamefont {Ellis},
  \citenamefont {Garcia}, \citenamefont {Nanopoulos},\ and\ \citenamefont
  {Olive}}]{Ellis:2015pla}%
  \BibitemOpen
  \bibfield  {author} {\bibinfo {author} {\bibfnamefont {J.}~\bibnamefont
  {Ellis}}, \bibinfo {author} {\bibfnamefont {M.~A.~G.}\ \bibnamefont
  {Garcia}}, \bibinfo {author} {\bibfnamefont {D.~V.}\ \bibnamefont
  {Nanopoulos}},\ and\ \bibinfo {author} {\bibfnamefont {K.~A.}\ \bibnamefont
  {Olive}},\ }\bibfield  {title} {\bibinfo {title} {{Calculations of Inflaton
  Decays and Reheating: with Applications to No-Scale Inflation Models}},\
  }\href {https://doi.org/10.1088/1475-7516/2015/07/050} {\bibfield  {journal}
  {\bibinfo  {journal} {JCAP}\ }\textbf {\bibinfo {volume} {07}},\ \bibinfo
  {pages} {050}},\ \Eprint {https://arxiv.org/abs/1505.06986} {arXiv:1505.06986
  [hep-ph]} \BibitemShut {NoStop}%
\bibitem [{\citenamefont {Ade}\ \emph {et~al.}(2018)\citenamefont {Ade} \emph
  {et~al.}}]{BICEP2:2018kqh}%
  \BibitemOpen
  \bibfield  {author} {\bibinfo {author} {\bibfnamefont {P.~A.~R.}\
  \bibnamefont {Ade}} \emph {et~al.} (\bibinfo {collaboration} {BICEP2, Keck
  Array}),\ }\bibfield  {title} {\bibinfo {title} {{BICEP2 / Keck Array x:
  Constraints on Primordial Gravitational Waves using Planck, WMAP, and New
  BICEP2/Keck Observations through the 2015 Season}},\ }\href
  {https://doi.org/10.1103/PhysRevLett.121.221301} {\bibfield  {journal}
  {\bibinfo  {journal} {Phys. Rev. Lett.}\ }\textbf {\bibinfo {volume} {121}},\
  \bibinfo {pages} {221301} (\bibinfo {year} {2018})},\ \Eprint
  {https://arxiv.org/abs/1810.05216} {arXiv:1810.05216 [astro-ph.CO]}
  \BibitemShut {NoStop}%
\bibitem [{\citenamefont {Greene}\ and\ \citenamefont
  {Kofman}(1999)}]{Greene:1998nh}%
  \BibitemOpen
  \bibfield  {author} {\bibinfo {author} {\bibfnamefont {P.~B.}\ \bibnamefont
  {Greene}}\ and\ \bibinfo {author} {\bibfnamefont {L.}~\bibnamefont
  {Kofman}},\ }\bibfield  {title} {\bibinfo {title} {{Preheating of
  fermions}},\ }\href {https://doi.org/10.1016/S0370-2693(99)00020-9}
  {\bibfield  {journal} {\bibinfo  {journal} {Phys. Lett. B}\ }\textbf
  {\bibinfo {volume} {448}},\ \bibinfo {pages} {6} (\bibinfo {year} {1999})},\
  \Eprint {https://arxiv.org/abs/hep-ph/9807339} {arXiv:hep-ph/9807339}
  \BibitemShut {NoStop}%
\bibitem [{\citenamefont {Datta}\ \emph {et~al.}(2023)\citenamefont {Datta},
  \citenamefont {Roshan},\ and\ \citenamefont {Sil}}]{Datta:2023pav}%
  \BibitemOpen
  \bibfield  {author} {\bibinfo {author} {\bibfnamefont {A.}~\bibnamefont
  {Datta}}, \bibinfo {author} {\bibfnamefont {R.}~\bibnamefont {Roshan}},\ and\
  \bibinfo {author} {\bibfnamefont {A.}~\bibnamefont {Sil}},\ }\bibfield
  {title} {\bibinfo {title} {{Flavor leptogenesis during the reheating era}},\
  }\href {https://doi.org/10.1103/PhysRevD.108.035029} {\bibfield  {journal}
  {\bibinfo  {journal} {Phys. Rev. D}\ }\textbf {\bibinfo {volume} {108}},\
  \bibinfo {pages} {035029} (\bibinfo {year} {2023})},\ \Eprint
  {https://arxiv.org/abs/2301.10791} {arXiv:2301.10791 [hep-ph]} \BibitemShut
  {NoStop}%
\bibitem [{\citenamefont {Buchmuller}\ \emph {et~al.}(2005)\citenamefont
  {Buchmuller}, \citenamefont {Di~Bari},\ and\ \citenamefont
  {Plumacher}}]{Buchmuller:2004nz}%
  \BibitemOpen
  \bibfield  {author} {\bibinfo {author} {\bibfnamefont {W.}~\bibnamefont
  {Buchmuller}}, \bibinfo {author} {\bibfnamefont {P.}~\bibnamefont
  {Di~Bari}},\ and\ \bibinfo {author} {\bibfnamefont {M.}~\bibnamefont
  {Plumacher}},\ }\bibfield  {title} {\bibinfo {title} {{Leptogenesis for
  pedestrians}},\ }\href {https://doi.org/10.1016/j.aop.2004.02.003} {\bibfield
   {journal} {\bibinfo  {journal} {Annals Phys.}\ }\textbf {\bibinfo {volume}
  {315}},\ \bibinfo {pages} {305} (\bibinfo {year} {2005})},\ \Eprint
  {https://arxiv.org/abs/hep-ph/0401240} {arXiv:hep-ph/0401240} \BibitemShut
  {NoStop}%
\bibitem [{\citenamefont {Casas}\ and\ \citenamefont
  {Ibarra}(2001)}]{Casas:2001sr}%
  \BibitemOpen
  \bibfield  {author} {\bibinfo {author} {\bibfnamefont {J.~A.}\ \bibnamefont
  {Casas}}\ and\ \bibinfo {author} {\bibfnamefont {A.}~\bibnamefont {Ibarra}},\
  }\bibfield  {title} {\bibinfo {title} {{Oscillating neutrinos and muon --->
  e, gamma}},\ }\href {https://doi.org/10.1016/S0550-3213(01)00475-8}
  {\bibfield  {journal} {\bibinfo  {journal} {Nucl. Phys.}\ }\textbf {\bibinfo
  {volume} {B618}},\ \bibinfo {pages} {171} (\bibinfo {year} {2001})},\ \Eprint
  {https://arxiv.org/abs/hep-ph/0103065} {arXiv:hep-ph/0103065 [hep-ph]}
  \BibitemShut {NoStop}%
\bibitem [{\citenamefont {Ashtari~Esfahani}\ \emph {et~al.}(2017)\citenamefont
  {Ashtari~Esfahani} \emph {et~al.}}]{Esfahani:2017dmu}%
  \BibitemOpen
  \bibfield  {author} {\bibinfo {author} {\bibfnamefont {A.}~\bibnamefont
  {Ashtari~Esfahani}} \emph {et~al.} (\bibinfo {collaboration} {Project 8}),\
  }\bibfield  {title} {\bibinfo {title} {{Determining the neutrino mass with
  cyclotron radiation emission spectroscopy\textemdash{}Project 8}},\ }\href
  {https://doi.org/10.1088/1361-6471/aa5b4f} {\bibfield  {journal} {\bibinfo
  {journal} {J. Phys. G}\ }\textbf {\bibinfo {volume} {44}},\ \bibinfo {pages}
  {054004} (\bibinfo {year} {2017})},\ \Eprint
  {https://arxiv.org/abs/1703.02037} {arXiv:1703.02037 [physics.ins-det]}
  \BibitemShut {NoStop}%
\bibitem [{\citenamefont {Esteban}\ \emph {et~al.}(2020)\citenamefont
  {Esteban}, \citenamefont {Gonzalez-Garcia}, \citenamefont {Maltoni},
  \citenamefont {Schwetz},\ and\ \citenamefont {Zhou}}]{Esteban:2020cvm}%
  \BibitemOpen
  \bibfield  {author} {\bibinfo {author} {\bibfnamefont {I.}~\bibnamefont
  {Esteban}}, \bibinfo {author} {\bibfnamefont {M.~C.}\ \bibnamefont
  {Gonzalez-Garcia}}, \bibinfo {author} {\bibfnamefont {M.}~\bibnamefont
  {Maltoni}}, \bibinfo {author} {\bibfnamefont {T.}~\bibnamefont {Schwetz}},\
  and\ \bibinfo {author} {\bibfnamefont {A.}~\bibnamefont {Zhou}},\ }\bibfield
  {title} {\bibinfo {title} {{The fate of hints: updated global analysis of
  three-flavor neutrino oscillations}},\ }\href
  {https://doi.org/10.1007/JHEP09(2020)178} {\bibfield  {journal} {\bibinfo
  {journal} {JHEP}\ }\textbf {\bibinfo {volume} {09}},\ \bibinfo {pages}
  {178}},\ \Eprint {https://arxiv.org/abs/2007.14792} {arXiv:2007.14792
  [hep-ph]} \BibitemShut {NoStop}%
\bibitem [{\citenamefont {Zyla}\ \emph {et~al.}(2020)\citenamefont {Zyla} \emph
  {et~al.}}]{Zyla:2020zbs}%
  \BibitemOpen
  \bibfield  {author} {\bibinfo {author} {\bibfnamefont {P.~A.}\ \bibnamefont
  {Zyla}} \emph {et~al.} (\bibinfo {collaboration} {Particle Data Group}),\
  }\bibfield  {title} {\bibinfo {title} {{Review of Particle Physics}},\ }\href
  {https://doi.org/10.1093/ptep/ptaa104} {\bibfield  {journal} {\bibinfo
  {journal} {PTEP}\ }\textbf {\bibinfo {volume} {2020}},\ \bibinfo {pages}
  {083C01} (\bibinfo {year} {2020})}\BibitemShut {NoStop}%
\bibitem [{\citenamefont {Antusch}\ \emph {et~al.}(2012)\citenamefont
  {Antusch}, \citenamefont {Di~Bari}, \citenamefont {Jones},\ and\
  \citenamefont {King}}]{Antusch:2011nz}%
  \BibitemOpen
  \bibfield  {author} {\bibinfo {author} {\bibfnamefont {S.}~\bibnamefont
  {Antusch}}, \bibinfo {author} {\bibfnamefont {P.}~\bibnamefont {Di~Bari}},
  \bibinfo {author} {\bibfnamefont {D.~A.}\ \bibnamefont {Jones}},\ and\
  \bibinfo {author} {\bibfnamefont {S.~F.}\ \bibnamefont {King}},\ }\bibfield
  {title} {\bibinfo {title} {{Leptogenesis in the Two Right-Handed Neutrino
  Model Revisited}},\ }\href {https://doi.org/10.1103/PhysRevD.86.023516}
  {\bibfield  {journal} {\bibinfo  {journal} {Phys. Rev. D}\ }\textbf {\bibinfo
  {volume} {86}},\ \bibinfo {pages} {023516} (\bibinfo {year} {2012})},\
  \Eprint {https://arxiv.org/abs/1107.6002} {arXiv:1107.6002 [hep-ph]}
  \BibitemShut {NoStop}%
\bibitem [{\citenamefont {Covi}\ \emph {et~al.}(1996)\citenamefont {Covi},
  \citenamefont {Roulet},\ and\ \citenamefont {Vissani}}]{Covi:1996wh}%
  \BibitemOpen
  \bibfield  {author} {\bibinfo {author} {\bibfnamefont {L.}~\bibnamefont
  {Covi}}, \bibinfo {author} {\bibfnamefont {E.}~\bibnamefont {Roulet}},\ and\
  \bibinfo {author} {\bibfnamefont {F.}~\bibnamefont {Vissani}},\ }\bibfield
  {title} {\bibinfo {title} {{CP violating decays in leptogenesis scenarios}},\
  }\href {https://doi.org/10.1016/0370-2693(96)00817-9} {\bibfield  {journal}
  {\bibinfo  {journal} {Phys. Lett. B}\ }\textbf {\bibinfo {volume} {384}},\
  \bibinfo {pages} {169} (\bibinfo {year} {1996})},\ \Eprint
  {https://arxiv.org/abs/hep-ph/9605319} {arXiv:hep-ph/9605319} \BibitemShut
  {NoStop}%
\bibitem [{\citenamefont {Harvey}\ and\ \citenamefont
  {Turner}(1990)}]{Harvey:1990qw}%
  \BibitemOpen
  \bibfield  {author} {\bibinfo {author} {\bibfnamefont {J.~A.}\ \bibnamefont
  {Harvey}}\ and\ \bibinfo {author} {\bibfnamefont {M.~S.}\ \bibnamefont
  {Turner}},\ }\bibfield  {title} {\bibinfo {title} {{Cosmological baryon and
  lepton number in the presence of electroweak fermion number violation}},\
  }\href {https://doi.org/10.1103/PhysRevD.42.3344} {\bibfield  {journal}
  {\bibinfo  {journal} {Phys. Rev. D}\ }\textbf {\bibinfo {volume} {42}},\
  \bibinfo {pages} {3344} (\bibinfo {year} {1990})}\BibitemShut {NoStop}%
\bibitem [{\citenamefont {Kofman}\ \emph {et~al.}(1994)\citenamefont {Kofman},
  \citenamefont {Linde},\ and\ \citenamefont {Starobinsky}}]{Kofman:1994rk}%
  \BibitemOpen
  \bibfield  {author} {\bibinfo {author} {\bibfnamefont {L.}~\bibnamefont
  {Kofman}}, \bibinfo {author} {\bibfnamefont {A.~D.}\ \bibnamefont {Linde}},\
  and\ \bibinfo {author} {\bibfnamefont {A.~A.}\ \bibnamefont {Starobinsky}},\
  }\bibfield  {title} {\bibinfo {title} {{Reheating after inflation}},\ }\href
  {https://doi.org/10.1103/PhysRevLett.73.3195} {\bibfield  {journal} {\bibinfo
   {journal} {Phys. Rev. Lett.}\ }\textbf {\bibinfo {volume} {73}},\ \bibinfo
  {pages} {3195} (\bibinfo {year} {1994})},\ \Eprint
  {https://arxiv.org/abs/hep-th/9405187} {arXiv:hep-th/9405187} \BibitemShut
  {NoStop}%
\bibitem [{\citenamefont {Garcia}\ \emph {et~al.}(2021)\citenamefont {Garcia},
  \citenamefont {Kaneta}, \citenamefont {Mambrini},\ and\ \citenamefont
  {Olive}}]{Garcia:2020wiy}%
  \BibitemOpen
  \bibfield  {author} {\bibinfo {author} {\bibfnamefont {M.~A.~G.}\
  \bibnamefont {Garcia}}, \bibinfo {author} {\bibfnamefont {K.}~\bibnamefont
  {Kaneta}}, \bibinfo {author} {\bibfnamefont {Y.}~\bibnamefont {Mambrini}},\
  and\ \bibinfo {author} {\bibfnamefont {K.~A.}\ \bibnamefont {Olive}},\
  }\bibfield  {title} {\bibinfo {title} {{Inflaton Oscillations and
  Post-Inflationary Reheating}},\ }\href
  {https://doi.org/10.1088/1475-7516/2021/04/012} {\bibfield  {journal}
  {\bibinfo  {journal} {JCAP}\ }\textbf {\bibinfo {volume} {04}},\ \bibinfo
  {pages} {012}},\ \Eprint {https://arxiv.org/abs/2012.10756} {arXiv:2012.10756
  [hep-ph]} \BibitemShut {NoStop}%
\bibitem [{\citenamefont {Moore}(1997)}]{Moore:1997im}%
  \BibitemOpen
  \bibfield  {author} {\bibinfo {author} {\bibfnamefont {G.~D.}\ \bibnamefont
  {Moore}},\ }\bibfield  {title} {\bibinfo {title} {{Computing the strong
  sphaleron rate}},\ }\href {https://doi.org/10.1016/S0370-2693(97)01046-0}
  {\bibfield  {journal} {\bibinfo  {journal} {Phys. Lett. B}\ }\textbf
  {\bibinfo {volume} {412}},\ \bibinfo {pages} {359} (\bibinfo {year}
  {1997})},\ \Eprint {https://arxiv.org/abs/hep-ph/9705248}
  {arXiv:hep-ph/9705248} \BibitemShut {NoStop}%
\end{thebibliography}%

			\clearpage
\newpage
\newpage
\maketitle
\onecolumngrid
\begin{center}
	\textbf{\large Effects of Reheating on Charged Lepton Yukawa Equilibration and Leptogenesis} \\ 
	\vspace{0.05in}
	{ \it \large Supplemental Material}\\ 
	\vspace{0.05in}
	{Arghyajit Datta, Rishav Roshan and Arunansu Sil}
\end{center}
\onecolumngrid
\setcounter{equation}{0}
\setcounter{figure}{0}
\setcounter{table}{0}
\setcounter{section}{0}
\setcounter{page}{1}
\makeatletter

This Supplemental Material is organized as follows. In Sec.~\ref{sup:1}, we discuss the equilibrium conditions for the SM interactions responsible for the production of lepton asymmetry. In Sec.~\ref{sup:2}, we analyze the dependence of $T_{\rm Max}$,  $T_{\rm RH}$ as well as the equilibration temperatures of the charged lepton Yukawa interactions on the effective individual inflaton-SM fermions coupling $y_{\phi}^f$. In Sec.~\ref{sup:3}, we provide the numerical analysis required to generate the baryon asymmetry in this modified setup and finally present the allowed parameter space in $y_{\phi}^f-M_1$ plane which 
is consistent with the observed baryon asymmetry.
\section{The equilibrium conditions}
\label{sup:1}
The Boltzmann equations (BE) describing the evolution of lepton asymmetry ($Y_{L}$) produced in leptogenesis are in general connected with different particle species and their asymmetries in a sense that various SM interactions, if found to be in chemical equilibrium, develop relations among number density asymmetries between different species. Such number density asymmetry ($n_{X} - n_{\bar{X}}$) for a particle species $X$ having number of degrees of freedom $g_X$ is connected to the respective chemical potential ($\mu_X$) via the relation \cite{Kolb:1990vq}
\begin{align}
	n_{X}-n_{\bar{X}}= \frac{g_X T^2}{6}
	\begin{cases}
		\mu_X & \text{for fermions}, \\
		2\mu_X & \text{for bosons},
	\end{cases} 
	\label{eq:chemical-pot}
\end{align}
as obtained from the respective distribution function. A particular SM interaction, whether or not in thermal equilibrium ($\equiv$ in chemical equilibrium) is determined by comparing the respective interaction rate ($\Gamma_{\rm{Int}}$) with Hubble expansion rate of the Universe ($\mathcal{H}$) at any temperature $T$. If $\Gamma_{\rm{Int}} \gtrsim \mathcal{H}$, the corresponding interaction rate is fast enough to establish an equilibrium situation such that the particles involved satisfy an algebraic relation among their chemical potentials. For example, if the charged lepton Yukawa interaction ($Y_{\alpha} {\bar{\ell}_{L_\alpha}} H e_{R_{\alpha}}$) of a particular flavor 
$\alpha$ satisfies such condition, it would maintain the relation
\begin{equation}
	\mu_{e_{R_\alpha}} - \mu_{\ell_{L_{\alpha}}} +\mu_H =0.
\end{equation}
Such inter-relations among the chemical potentials are important to quantify the lepton asymmetry we are interested in, as explained below. \\

The lepton asymmetry in terms of number densities of leptons and anti-leptons, normalized to the entropy density ($s$), is defined by 
\begin{equation}
	Y_L (\equiv Y_{\Delta_L}) = \frac{n_{\Delta_L}}{s} = \sum_\alpha \frac{1}{s}\Big[ \left(n_{e_{L_\alpha}}-n_{\bar{e}_{L_\alpha}}\right)+\left(n_{\nu_{L_\alpha}}-n_{\bar{\nu}_{L_\alpha}}\right)+\left(n_{{e_R}_\alpha}-n_{\bar{e}_{R_\alpha}}\right)\Big], \label{Y_L}\\
\end{equation}
where both the $SU(2)_L$ doublet and singlet leptons are included for general purpose. Using Eq. \eqref{eq:chemical-pot} and the fact that the two members of a $SU(2)_L$ multiplet will have same chemical potential, $i.e. ~\mu_{\ell}\equiv \mu_{e_L}=\mu_{\nu_L}$ resulting from the vanishing chemical potential of the electroweak gauge bosons above the electroweak scale, when the corresponding gauge interaction is in equilibrium, the total lepton number asymmetry can be expressed as:
\begin{align}
	Y_L &= \frac{T^2}{6s}\sum_\alpha \left( 2 \mu_{\ell_\alpha}+ \mu_{e_{R_\alpha}}\right)		\label{eq:Y-mu},\\
	& \equiv
	\sum_\alpha \left(2 Y_{\Delta_{\ell_\alpha}} + Y_{\Delta_{e_{R_\alpha}}}\right).
	\label{eq-L-2}
\end{align}

Now to determine whether few or all of these chemical potentials $\mu_{\ell_\alpha}, \mu_{e_{R_\alpha}}$, are non-zero and the relationships these chemical potentials might carry with other non-zero chemical potentials, one needs to identify the temperature regime ($T \sim M_1$) where the lepton asymmetry is being generated from the decay of the lightest right handed neutrino (RHN) of mass $M_1$. Then, following conditions on the chemical potentials for different SM fields arise as a result of corresponding SM interactions, attaining thermal equilibrium with the decrease in the temperature of the Universe  (in a radiation dominated era) should be considered. 

\begin{itemize}

	\item For temperature above $10^{13}$ GeV, interactions mediated by the SM gauge couplings and the top quark Yukawa coupling comes into equilibrium. Consequently, the following conditions on the chemical potentials of the participating particles arise: 
	(i) The chemical potential of the gauge bosons (including gluons, $W^i_{\mu}$ and $B_\mu$) are zero, $i.e.,$ 
	\begin{equation}
		\mu_g = \mu_{W^i}=\mu_{B}=0, 
	\end{equation}
	with $i = 1,2,3.$
	As a result, two members of a $SU(2)_L$ doublet reach chemical equilibrium and have same chemical potential: $\mu_Q\equiv \mu_{u_L}=\mu_{d_L}, \quad \mu_{\ell}\equiv \mu_{e_L}=\mu_{\nu_L}, \quad \mu_H\equiv\mu_{H^+}=\mu_{H^0}$. This is already incorporated in Eq. \eqref{eq:Y-mu}. 
	(ii) Furthermore, equilibration of top quark Yukawa interaction yields, 
	\begin{equation}
		\mu_t=\mu_{Q_3}+\mu_H.
	\end{equation}
	
	\item A temperature independent constraint follows from the hypercharge neutrality condition, leading to 
	\begin{align}
		\sum_\alpha (\mu_{Q_\alpha}+2 \mu_{u_\alpha}-\mu_{d_\alpha}-\mu_{\ell_\alpha}-\mu_{e_{R_\alpha}})+2\mu_{H}=0,
		\label{eq-hy}
	\end{align}
	where $u_\alpha,d_\alpha$ represents the $SU(2)_L$ singlet quarks of $\alpha$ flavor respectively.

	\item Around $T\sim 10^{13}$ GeV, (i) strong sphalerons come to equilibrium~\cite{Moore:1997im,Bento:2003jv} leading to:
	\begin{align}
		\sum_\alpha(2\mu_{Q_\alpha}-\mu_{u_\alpha}-\mu_{d_\alpha})=0.
	\end{align}
	(ii) On the other hand, equilibrium temperature of electroweak sphalerons is found to be  smaller than that of strong sphaleron equilibrium temperature~\cite{Arnold:1996dy,Bodeker:1998hm,Arnold:1998cy,Bento:2003jv} by an order of magnitude. Once in equilibrium, this enforces:
	\begin{align}
		\sum_\alpha (3\mu_{Q_\alpha}+\mu_{\ell_\alpha})=0.
	\end{align} 
	(iii) At this stage, all the charged lepton Yukawa interactions are out of equilibrium and hence, the respective chemical potentials of the $SU(2)_L$ singlet right handed charged leptons can be set to zero ($i.e. \sum_\alpha Y_{\Delta_{e_{R_\alpha}}} = 0$). Then the lepton asymmetry, if produced from $N_1$ decay in this regime, along lepton doublet direction only contribute to the total lepton asymmetry and becomes, 
	\begin{align}
		Y_L= \sum_\alpha Y_{\Delta_{\ell_\alpha}}.
	\end{align} 
	
	\item However, when the temperature further falls down to $T < 10^{13}$ GeV, charged lepton (and remaining quark Yukawa interactions) Yukawa interactions start entering equilibrium one after other. As the temperature drops, interactions with larger Yukawa coupling attains thermal equilibrium faster. Once they are in equilibrium, the following relations among chemical potentials:
	\begin{align}
		\mu_{e_{R_\alpha}}-\mu_{\ell_\alpha}+\mu_H&=0 \quad \text{for~~charged~leptons},\\
		\mu_{u_\alpha}-\mu_{Q_\alpha}-\mu_H&=0 \quad \text{for~up-type~quarks},\\
		\mu_{d_\alpha}-\mu_{Q_\alpha}+\mu_H&=0\quad \text{for~down-type~quarks}
		\label{eq:chemicalyukawa}
	\end{align}
	can be set. It is found that while $Y_{\tau}$ interaction enters equilibrium at $T = 5 \times 10^{11}$ GeV, that of $Y_{\mu}$ and $Y_{e}$ interactions are at $10^9$ GeV and $5 \times 10^4$ GeV respectively. In this case, depending upon which charged lepton Yukawa interactions are in equilibrium, following Eq. \eqref{eq-L-2}, contributions from $Y_{\Delta_{e_{R_\alpha}}}$ should also be incorporated in evaluating $Y_{L}$. 
	
\end{itemize}

Finally, to quantify the lepton asymmetry contribution coming from right handed leptons or from the lepton doublets, one needs take into account the chemical potential relations (and other constraints) from Eq.~\eqref{eq-hy}-\eqref{eq:chemicalyukawa}, which eventually leads to the $C^\ell, C^H$ matrices mentioned in Eq[10] of the main text (see\cite{Barbieri:1999ma,Nardi:2005hs,Nardi:2006fx} for details). 
Note that the above discussion, particularly the temperature regimes where different SM interactions enter in equilibrium, is based on the assumption that the Universe undergoes a radiation dominated era during such decouplings. However, as evaluated in the text, the regime $T_{\rm{Max}} > T (\sim M_1)  > T_{\rm{RH}}$ turns out to be an important one where the charged lepton Yukawa interactions are greatly affected and the respective equilibration conditions are to be applied with care.

\section{Estimating $T_{\rm{Max}}, T_{\rm{RH}}$ and Equilibration Temperatures  as function of $y_{\phi}^f$}	
\label{sup:2}

In this section, we evaluate the quantities important for studying reheating. In our work, the process of reheating proceeds perturbatively through the effective inflaton-SM fermion interaction $y_\phi^f \phi \bar{f}f$.  As discussed in the main paper, this 
effective individual coupling $y_\phi^f$ plays a crucial role in understanding the entire reheating process by providing estimates for $T_{\rm{Max}}$ and $T_{\rm{RH}}$. The respective equilibration temperatures (ET) of different charged lepton Yukawa interactions depend crucially on the temperature profile, dictated by $y_{\phi}^f$, during this extended reheating period which in turn affects the flavor effect on leptogenesis. This makes $y_\phi^f$ an important parameter in the present setup. Though in the text, we consider a specific choice of $y_{\phi}^f$ for the demonstration purpose, it is indeed intriguing to have analytic expressions of the quantities like $T_{\rm{Max}}, T_{\rm{RH}}$ and ET as function of $y_{\phi}^f$, for general purpose. This will also remain helpful for scanning the parameter space of the work as investigated in the next section.  

\subsection{Estimating $T_{\text{Max}}$ and $T_{\text{RH}}$}

To understand the reheating period, we first study the evolutions of inflaton and radiation density via Eq. 4 of the text as below, 
\begin{align}
	&\frac{d\rho_{\phi}}{dt}+3 \left(\frac{3n}{n+2}\right) \mathcal{H} \rho_{\phi}=-\Gamma_\phi  \rho_\phi,\label{eq:a1}\\
	&\frac{d\rho_R}{dt}+4 \mathcal{H} \rho_R= \Gamma_{\phi}\rho_\phi, \label{eq:a2}
\end{align}	
where the $y_{\phi}^f$ involvement comes through the decay width $\Gamma_\phi= g_f^2\frac{{y_\phi^f}^2}{8\pi} m_\phi$ with $g_f$ representing the effective degrees of freedom (considering different generations of leptons and quarks of different colors) and $n$ corresponds to the power involved in the inflationary potential. Here $m_\phi$ is the effective mass of the inflaton and in the adiabatic approximation, it can be estimated as:
\begin{align}
	m_\phi= \partial_\phi^2 V(\phi)= \lambda^{2/n}n(n-1) \rho_\phi^{\frac{n-2}{n}}M_P^{\frac{2(4-n)}{n}}.
	\label{eq:a3}
\end{align}
In the early Universe just after inflation, the decay rate of the inflaton can be safely assumed to be smaller than the expansion rate of the Universe. This helps to solve Eq.~\eqref{eq:a1} analytically (by neglecting the right-hand side of the equation) and is given by
\begin{align}
	\rho_\phi(a) \simeq \rho_{\rm end} \left(\frac{a}{a_{\rm end}}\right)^{-6n/(n+2)}, 
	\label{eq:a4}
\end{align}
where $\rho_{\rm end }$ represents the energy density of the inflaton at the end of the inflation and can be estimated by setting the slow roll parameter $\epsilon=1$ which leads to $\rho_{\rm end}=\frac32 V(\phi_{\rm end})$\cite{Ellis:2015pla} (for details, see T-attractor inflation models \cite{Kallosh:2013hoa, Garcia:2020eof}). 
The radiation energy density $\rho_{\phi}$ can also be estimated by solving the Eq.~\eqref{eq:a2} after plugging Eqs. \eqref{eq:a3} and \eqref{eq:a4} into it. This gives,
\begin{align}
	\rho_R \simeq \frac{g_f^2 {y_\phi^f}^2}{8\pi} \rho_{\rm end}^{\frac{n-1}{n}} \sqrt{n(n-1)}\left(\frac{n+2}{14-2n}\right)\lambda^{1/n} M_P^{4/n}
	\left(\frac{a_{\rm end}}{a}\right)^4\left\{\left( \frac{a}{a_{\rm end}}\right)^{\frac{14-2n}{n+2}}-1\right\}.
	\label{eq:a5}
\end{align}

Note that, irrespective of the choice of $n$, the radiation energy density depends on the effective coupling as $\rho_R\propto {y_\phi^f}^2$. As a result, increasing $y_\phi^f$ increases the energy density of the radiation. Now, assuming instantaneous equilibration of the decay products of inflaton, the temperature of the Universe can be easily obtained from $\rho_R$ using
\begin{align}
	T= \left(\frac{30 \rho_R}{\pi^2 g_*}\right)^\frac14,
	\label{eq:a6}
\end{align}
where $g_*$ is the effective relativistic degrees of freedom. Hence, just after inflation, the temperature of the Universe increases to a maximum value $T_{\rm Max}$ that can be obtained by maximizing $T$ of Eq.~\eqref{eq:a6} against the scale factor $a$ with the help of Eq. \eqref{eq:a5},
\begin{align}
	T_{\rm Max}=\left[\frac{15 g_f^2 {y_\phi^f}^2}{16 g_* \pi^3}\rho_{\rm end}^{\frac{n-1}{n}} \sqrt{n(n-1)} \left(\frac{3n-3}{2n-4}\right)^{\frac{3(n-1)}{7-n}} M_P^{4/n}\right]^{1/4}.
	\label{eq:a7}
\end{align}

\begin{figure}[t]
	\centering
	{\includegraphics[width=0.7\linewidth]{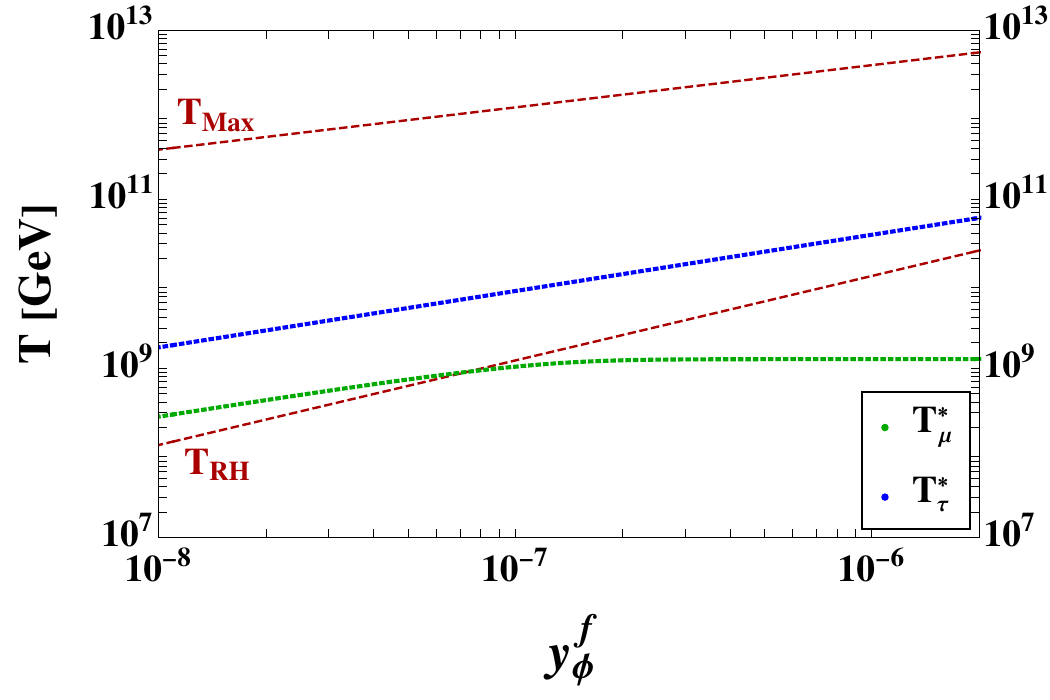}}
	\caption{Variation of $T_{\rm Max}$, $T_{\rm RH}$ and ETs with respect to $y_\phi^f$: The upper (lower) brown dashed line represents the variation of $T_{\rm Max}$ ($T_{\rm RH}$) while $T^*_\tau, T^*_\mu,$ are represented by blue and green dotted lines respectively.}
	\label{fig:a1_sm}
\end{figure}

\noindent Once $T_{\rm Max}$ is reached, the temperature profile starts falling as $T\propto a^{-\frac{3n-3}{2n+4}}$ and its continues to fall till an equality between matter and radiation energy density is reached. This point, known as the $T_{\rm RH}$, can be estimated by comparing $\rho_\phi= \rho_R$. Using Eqs. \eqref{eq:a4} and \eqref{eq:a5}, one obtains

\begin{align}
	T_{\rm RH}= \left[ (g_f y_\phi^f)^{2n} 
	\frac{15\{3n(n-1)\}^{n/2} \lambda}{2^{4n-1}\pi^{2+n}g_*}
	\left(\frac{n+2}{7-n}\right)^n M_P^4
	\right]^{1/4}.
	\label{eq:a8}
\end{align}
Following Eq.\eqref{eq:a7} and \eqref{eq:a8}, one finds that $T_{\rm Max}\propto {y_\phi^f}^{1/2}$  while  $T_{\rm RH}\propto {y_\phi^f}$ (for $n=2$). Such dependence becomes explicit in Fig. \ref{fig:a1_sm} where we plot $T_{\rm Max}$ (upper dashed brown line) and $T_{\rm{RH}} $ (bottom dashed brown line) as function of $y_{\phi}^f$. We restrict ourselves with the choice of $y_{\phi}^f < 3.3 \times 10^{-6}$ as beyond this value, explosive production of fermions via preheating may take place, which we intend to avoid in the present work.

\subsection{Equilibrium Temperature of $e_{R_{\alpha}}$ as function of $y_\phi^f$}

Next, we aim to identify the dependence of equilibration temperature of different charged lepton Yukawa interaction on $y_\phi^f$. In order to estimate that, one may equate
\begin{align}
	\langle \Gamma_\alpha\rangle= \mathcal{H}, 
	\label{eq:gbyh}
\end{align}
with $\langle \Gamma_\alpha\rangle= \frac{\pi Y^2_{\alpha}}{192 \zeta(3)} \frac{m^2_h (T)}{T}$ associated to a specific Yukawa interaction $Y_{\alpha}$. During the extended reheating era defined by $T_{\text{Max}} > T > T_{\text{RH}}$, due to the $\rho_{\phi}$ dominance, the Hubble can be approximated effectively by $\mathcal{H}^2\simeq \frac{\rho_\phi}{3 M_p^2}$. Thereafter, using $m_h(T)\simeq0.6T$ in $\langle \Gamma_\alpha\rangle$ and Eq.~\eqref{eq:a4} into Eq.~\eqref{eq:gbyh}, we write, 
\bea
\frac{\pi Y^2_{\alpha}}{192 \zeta(3)} \frac{(0.6T)^2}{T} &\simeq& \sqrt{\frac{\rho_{\rm end}}{3M_p^2} \left(\frac{a}{a_{\rm end}}\right)^{-6n/(n+2)}}.
\label{eq:gbyh2}
\eea
Further, in Eq.~\eqref{eq:gbyh2} one can replace the $a/a_{\rm end}$ dependence in terms of $T$ using Eq.~\eqref{eq:a5} and \eqref{eq:a6} so as to obtain an analytical expression of ET of respective charged lepton Yukawa interaction. At a much later stage (in the limit $a \gg a_{\rm end}$), the modified ET associated to $Y_{\alpha}$ in interaction $T^*_{\alpha}$ can then be written as,
\begin{align}
	T^*_{\alpha}= \mathbb{C}^{\frac{n-1}{n+1}} Y_\alpha^{\frac{2(n-1)}{n+1}} (g_f y_\phi^f)^{\frac{n}{n+1}},
	\label{eq:equib}
\end{align}
where $\mathbb{C}$ is a quantity independent of $y_\phi^f$, given by
\begin{align}
	\mathbb{C}=5 \sqrt{3} \times 10^{-3} 
	\lambda^{\frac{1}{2(n-1)}} M_p^{\frac{n+1}{n-1}}
	\Big\{3n(n-1)\Big\}^{\frac{n}{4(n-1)}}
	\left(\frac{n+2}{14-2n}\right)^{\frac{n}{2(n-1)}} 
	\left(\frac{30}{8 \pi^3 g_*}\right)^{\frac{n}{2(n-1)}}.
\end{align}
Such analytical estimate of the ETs as a function of $y_{\phi}^f$ during the extended period of reheating ($T^*_{\alpha} \propto {y_{\phi}^f}^{2/3}$ with $n = 2$) via Eq.~\eqref{eq:equib} 
is in a good agreement with the numerical results we had in the main paper. The ETs for $\tau_R$ and $\mu_R$ are plotted in 
Fig.\ref{fig:a1_sm} as a function of $y_{\phi}^f$. Note that, $e_R$ does not equilibrate above the $T_{\rm RH}$ obtained, within the specified range of $y_{\phi}^f$ considered here which means $T^*_{e} = T^*_{0(e)}$ beyond this point (in the radiation dominated era). 
As expected, both $T^*_\tau$ (blue dotted line) and  $T^*_\mu$ (green dotted line) increase with the increase in $y_{\phi}^f$ during this extended reheating period. At some point with a typical value of $y_{\phi}^f$, $T^*_\mu$ coincides with $T_{\rm{RH}}$ and thereafter $T^*_{\mu}$ saturates to its value $T^*_{0\mu}$ as in the radiation dominated era (indicated by the constant line). However, the $Y_{\tau}$ interaction continues to have modified ET throughout the entire range of $y_{\phi}^f$ considered.

At this stage, one may wonder what happens to the other SM interactions like the top, bottom, charm Yukawa interactions, EW sphaleron interactions, etc. In this regard, the point is to note that the source of the modified ET related to any interaction is related to $\mathcal{H}$ and hence all these other interactions should also be affected similarly. So overall a scaling effect $w.r.t.$ the charged lepton Yukawa ET should be observed. As the strength of these interactions is more compared to $Y_{\tau}$, it is natural to expect these interactions to be in equilibrium by the time $T^*_{\tau}$ has reached.

\section{Lepton asymmetry as function of $M_1$ and $y_\phi^f$}
\label{sup:3}
\begin{figure}[t]
	\centering
	{\includegraphics[width=0.7\linewidth]{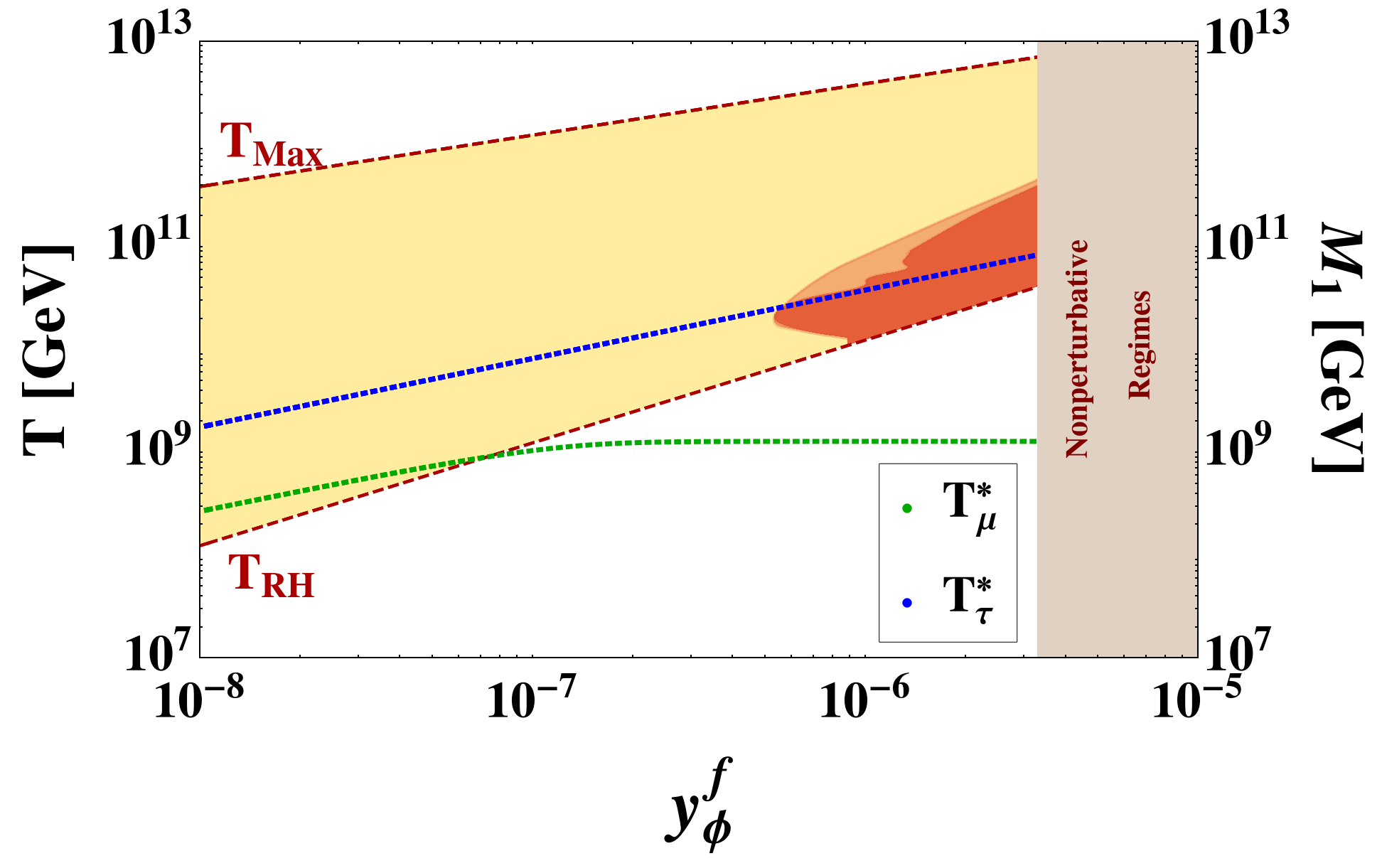}}
	\caption{ Dark brown patch represents the allowed value of RHN mass $M_1$ and coupling $y_\phi^f$ which can produce the correct baryon asymmetry for the scenario with modified flavor regimes. On the contrary, light brown patch represents the allowed region when change in flavor effect was not taken into account.  Here $y_\phi^f>3.3 \times 10^{-5}$ indicates the nonperturbative regime.}
	\label{fig:a2_sm}
\end{figure}

With the above understanding on the $T_{\rm{Max}}$ and $T_{\rm{RH}}$ as a function of $y_{\phi}^f$, it is evident that the parameter space for the lightest RHN of mass $M_1$ for case (b) of the main paper ($T_{\rm{Max}} > M_1 > T_{\rm{RH}}$) is limited within the region bounded by the $T_{\rm{Max}}$ and $T_{\rm{RH}}$ lines of Fig. \ref{fig:a1_sm}. As a result, for the analysis of the lepton asymmetry generation from the decay of RHN $N_1$, we focus primarily on this range of $M_1$, 
indicated by the yellow shaded region of Fig. \ref{fig:a2_sm} corresponding to $ y_{\phi}^f < 3.3 \times 10^{-6}$. 
Now, the lightest RHN mass $M_1$ being smaller than the temperature of the bath for the period: $T_{\rm{Max}} - T(\sim M_1)$ (for a specific $y_{\phi}^f$), it should be allowed to be produced via inverse decay. Furthermore, for the rest of the reheating period, $i.e.$ for: $T(\sim M_1) - T_{\rm{RH}}$, its decay to the SM bath particles also contributes to the radiation. Hence, in addition to $\rho_{\phi}$ and $\rho_R$, contribution of $\rho_{N_1}$ needs to be incorporated to study the evolution of different components and final lepton asymmetry. To estimate such evolutions, we involve a set of coupled Boltzmann equations as, 
\besub
\bea
\frac{d\rho_\phi }{dt}+3\mathcal{H}\rho_\phi&=&- \Gamma_\phi \rho_\phi ,\label{eq:phi2}\\ 
\frac{d\rho_R }{dt}+4\mathcal{H}\rho_R&=&  \Gamma_\phi \rho_\phi+ \langle\Gamma_{N_1}\rangle (\rho_{N_1}-\rho_{N_1}^{\text{eq}}), \label{eq:r2}\\
\frac{d \rho_{N_1}}{dt}+3\mathcal{H}\rho_{N_1}&=& -\langle\Gamma_{N_1}\rangle (\rho_{N_1}-\rho_{N_1}^{\text{eq}}),
\label{eq:n2}\\
\frac{d n_{\Delta}}{dt}+3\mathcal{H} n_{\Delta}&=&-\langle\Gamma_{N_1}\rangle \left[\frac{\varepsilon_\ell}{M_1}(\rho_{N_1}-\rho_{N_1}^{\rm{eq}})+\frac{n_{N_1}^{\rm{eq}}}{2 n_\ell^{\rm{eq}}}n_{\Delta}\right],
\label{eq:b-l2}
\eea
\label{BE_all}
\eesub
where we consider the case with $n=2, \mathcal{H} = \sqrt{\frac{\rho_{\phi}+ \rho_{R} +\rho_{N_1}}{3M^2_P}}$ and $n_{\Delta} = n_{B-L} \equiv - n_{\Delta_L}$ as defined in Eq. \ref{Y_L}. For solution purpose, it is found to be convenient to replace the $t$-dependance in terms of scale factor $a$ using the relation $\frac{d}{da}\equiv \frac{1}{a \mathcal{H}} \frac{d}{dt}$. The transformed equations are mentioned in the main paper. 

At this stage, we recollect that apart from $M_1$ and $y_{\phi}^f$, the lepton asymmetry, generated as a result of $N_1$ decay, 
is also controlled by the neutrino Yukawa coupling matrix $Y_\nu$, via $\Gamma_{N_1}$ in the BEs above. Introducing CI-parametrisation makes $Y_\nu$ a function of $M_1$, Re[$\theta_R$] and Im[$\theta_R$], provided the heavier RHN masses carry a specific ratio with $M_1$ together with the neutrino oscillation parameters fixed from experimental data (see the main text for the details). The inflation related parameters such as $m_{\phi}$ and hence $\Gamma_{\phi}$ are so evaluated as to satisfy the inflationary predictions in terms of spectral index $n_s$ and tensor to scalar ratio $r$, as stated in the main paper. With this understanding, we now proceed below to discuss the modification expected on flavor leptogenesis as a result of shift(s) in the ET(s) during extended reheating period and thereafter provide a guideline for scanning the parameter space that could correctly produce the observed baryon asymmetry. 

While solving the above BEs to evaluate the lepton asymmetry, the effect of individual lepton flavor interactions enters in the picture if one of the charged lepton Yukawa interactions, say $Y_{\tau}$, becomes comparable to the neutrino Yukawa interaction. As the RHN $N_1$ stays out of equilibrium during its decay for lepton asymmetry production, checking of equilibration condition for tau Yukawa interaction is sufficient to identify whether these interactions are strong enough. In such a situation, lepton (anti-lepton) states produced from the RHN decay start interacting with right-handed tau leptons and as a result, asymmetry produced in the tau flavor direction gets enhanced or diluted differently compared to other flavor directions. This can impact the total lepton asymmetry. Hence, one needs to take into account the time evolution of the lepton asymmetry along the tau flavor state and a coherent superposition state of muon and electron separately which is the core of flavor leptogenesis. On the other hand, at a lower temperature, when $\mu$-Yukawa interaction becomes dominant over the expansion rate of the Universe, the quantum coherence of the lepton states produced from the RHN decay breaks down completely and one needs to look for the evolution of the lepton asymmetry along $e,~\mu$ and $\tau$ separately. The BEs that describe such scenarios is given by (the Eq. [10] of the main paper),
\begin{align} 
	\frac{dn_{\Delta_\alpha}}{dt}+ 3 \mathcal{H}n_{\Delta_\alpha}=-\langle\Gamma_{N_1}\rangle \left[\frac{\varepsilon_{\ell_\alpha}}{M_1}(\rho_{N_1}-\rho_{N_1}^{\text{eq}})
	+\frac{1}{2}K^0_\alpha\sum_\beta(C^\ell_{\alpha \beta}+C^H_{\beta})\frac{n_{N_1}^{\rm{eq}}}{n_\ell^{\rm{eq}}}n_{\Delta_\beta}\right],
	\label{eq:yblij}
\end{align}
where $\alpha=e,\kappa$ when only $Y_\tau$ is in thermal equilibrium, while $\alpha=e,\mu,\tau$ when both $Y_\tau$ and $Y_\mu$ become non negligible. Here, $\varepsilon_{\ell_\alpha}$ denotes the CP asymmetry produced in the relevant lepton flavor directions, while $K^0_\alpha$ represents the ratio between tree level decay rate of lightest RHN to relevant lepton flavor $\alpha$ and total decay rate of $N_1$. The asymmetry produced in the Higgs and lepton doublet sector is converted to $B/3-L_\alpha$ number asymmetry via $C^\ell$ and $C^H$ matrices. 

We have already noticed that the parameters controlling the lepton asymmetry production are essentially: $y_{\phi}^f, M_1$, Re[$\theta_R$] and Im[$\theta_R$]. Among these, $y_{\phi}^f$ is bounded from above by $3.3 \times 10^{-6}$ as stated before while $M_1$ is constraint to be between $T_{\rm{Max}}$ and $T_{\rm{RH}}$ (for a specific $y_{\phi}^f$) as indicated by the yellow patch of Fig. \ref{fig:a2_sm} for case (b) (the primary focus of the main paper). Keeping these in mind, we solve the BEs (Eq.~\eqref{eq:phi2}-\eqref{eq:n2} along with Eq.~\eqref{eq:b-l2} or Eq.~\eqref{eq:yblij} depending on the flavor regimes) for $y_{\phi}^f < 3.3 \times 10^{-6}$ and $M_1$ being scanned across the yellow region of Fig. \ref{fig:a2_sm} while allow Re[$\theta_R$] and Im[$\theta_R$] to vary arbitrarily with the sole constraint that such values must not make any element of $Y_{\nu}$ non-perturbative $i.e.~ (Y_{\nu})_{ij} < \mathcal{O}(1)$. 

It is found that only a stipulated parameter space, as indicated by the dark brown patch in Fig.~\ref{fig:a2_sm}, satisfies the correct amount of final baryon asymmetry. The information on the RHN neutrino mass $M_1$ is indicated in the right side $Y$-axis. Note that, this parameter space is inclusive of the shift in the equilibration temperature(s) 
of the charged lepton Yukawa interaction(s) that we observe in this work. To make a comparison, we include the light brown patch as well which corresponds to the parameter space producing correct amount of final baryon asymmetry production, however, without taking into account the shift in the ETs during this extended era of reheating. We notice that the parameter space is modified and more constrained when the impact of shift in ETs is correctly embedded in flavor leptogenesis scenario. 
In the main paper, we discuss such impact on lepton asymmetry production, due to modified (smaller compared to standard case) ETs, with the choice of a specific benchmark point due to the limited space. Here however, we scan the entire parameter space which in fact provides a clear confirmation about the importance of re-evaluating the charged lepton Yukawa equilibration during the reheating phase when lepton asymmetry production is simultaneously taking place. 

We further notice that the blue dashed line, representative of the ET of tau leptons $T^*_{\tau}$, goes through the dark brown patch and marks a demarkation between the unflavored and two flavored regimes of thermal leptogenesis. Note that for $M_1$ values falling above this line, $N_1$ decay effectively contributes to lepton asymmetry production while none of the charged lepton Yukawa interaction enters in equilibrium and as a result this case correspond to unflavored regime of leptogenesis. Contrary to this, for $M_1$ values lying below this line, $N_1$ decay starts after the $Y_{\tau}$ interaction equilibrates so as to consider $\tau$ and $\kappa$ (a linear combination of $e$ and $\mu$) flavor directions separately as 
for such $M_1$, the $T^*_{\mu}$ remains below the $T_{\rm RH}$. Hence, these ($M_1$ below the blue dashed line) points belong to two flavor regime of thermal leptogenesis. Looking at this plot, we conclude that successful leptogenesis can be achieved with a minimum value of $T\sim M_1=1.3 \times 10^{10}$ GeV and $T_{\rm RH}=1.1 \times 10^{10}$ GeV. At this stage, it is also pertinent to mention that the energy density of the RHN always remains subdominant to the radiation energy density. This is 
due to the fact that the RHN $N_1$ exists in the Universe for a very short period of time as it has to be produced and decay in between the era of $T_{\text{max}}$ and $T_{\text{RH}}$. As a result of which, the reheating temperature remains effectively unchanged even in the presence of RHN. This behavior is also depicted in Fig. 3 of the main text. A deviation of such a situation is discussed in a follow-up paper \cite{Datta:2023pav}.

\end{document}